%% file: master.tex
\documentstyle[12pt,fleqn,epsfig,fullpage]{article}
\newcommand{\be}{\begin{eqnarray}}
\newcommand{\ee}{\end{eqnarray}}

\begin{document}
\title{Event-by-Event Fluctuations in Heavy Ion Collisions and
the QCD Critical Point}

\author{M. Stephanov${}^{(a)}$, K. Rajagopal${}^{(b)}$ and  
E. Shuryak${}^{(c)}$\\[0.5ex]
{\normalsize
${}^{(a)}$ Institute for Theoretical Physics,}\\ 
{\normalsize State University of New York, 
Stony Brook, NY 11794-3840}\\[0.5ex]
{\normalsize ${}^{(b)}$ Center for Theoretical Physics,}\\ 
{\normalsize Massachusetts Institute of Technology,
Cambridge, MA 02139}\\[0.5ex]
{\normalsize ${}^{(c)}$ Department of Physics and Astronomy,}\\ 
{\normalsize State University of New York, 
     Stony Brook, NY 11794-3800}
}

\date{\small 
March 8 (rev. July 8), 1999; ITP-SB-99-4, MIT-CTP-2834, SUNY-NTG-99-3}

\maketitle

\begin{abstract} 

The event-by-event fluctuations of suitably chosen observables in
heavy ion collisions at SPS, RHIC and LHC can tell us about the
thermodynamic properties of the hadronic system at freeze-out.  By
studying these fluctuations as a function of varying control
parameters, it is possible to learn much about the phase diagram of
QCD. As a timely example, we stress the methods by which present
experiments at the CERN SPS can locate the second-order critical
endpoint of the first-order transition between quark-gluon plasma and
hadron matter.  Those event-by-event signatures which are
characteristic of freeze-out in the vicinity of the critical point
will exhibit nonmonotonic dependence on control parameters.  We focus
on observables constructed from the multiplicity and transverse
momenta of charged pions.  We first consider how the event-by-event
fluctuations of such observables are affected by Bose-Einstein
correlations, by resonances which decay after freeze-out and by
fluctuations in the transverse flow velocity.  We compare our
thermodynamic predictions for such noncritical event-by-event
fluctuations with NA49 data, finding broad agreement.  We then focus
on effects due to thermal contact between the observed pions and a
heat bath with a given (possibly singular) specific heat, and due to
the direct coupling between the critical fluctuations of the sigma
field and the observed pions.  We also discuss the effect of the pions
produced in the decay of sigma particles just above threshold after
freeze-out on the inclusive pion spectrum and on multiplicity
fluctuations.  We estimate the size of these nonmonotonic effects
which appear near the critical point, including restrictions imposed
by finite size and finite time, and conclude that they should be
easily observable.

\end{abstract}


\input{intro}

\input{chap2}

\input{chap3}

\input{chap4}

\input{chap5}

\input{chap6}

\input{sum}

\input{appendix}

\input{bib}
\end{document}

%% file: intro.tex
\section{Introduction and Outline}

The goal of this paper is to motivate a program of heavy ion collision
experiments aimed at discovering an important qualitative feature of
the QCD phase diagram, namely the critical endpoint at which a line of
first order phase transitions separating quark-gluon plasma from
hadronic matter comes to an end~\cite{SRS}. The possible existence of
such an end point, denoted E, in the temperature ($T$) vs. baryon chemical
potential ($\mu$) plane has recently been emphasized and its universal
critical properties have been described~\cite{BeRa97,HaJa97}. The
point E can be thought of as a descendant of a tricritical point in
the phase diagram for 2-flavor QCD with {\it massless} quarks.  As
pointed out in~\cite{SRS}, observation of the signatures of freezeout
near E would confirm that heavy ion collisions are probing above the
chiral transition region in the phase diagram.  Furthermore, we would
learn much about the qualitative landscape of the QCD phase diagram.
\par
 In a previous letter~\cite{SRS}, we have laid
out the basic ideas for observing the critical endpoint.
The signatures 
proposed in~\cite{SRS} are based on the fact that such a point is a
genuine thermodynamic singularity at which 
susceptibilities diverge and the order parameter
fluctuates on long wavelengths. The resulting
signatures all share one common property: they are {\em
nonmonotonic\/} as a function of an experimentally varied parameter
such as the collision energy, centrality, rapidity or 
ion size. The goal of the present paper is to develop a set of tools which 
will allow heavy ion collision experiments to discover the critical
endpoint through the analysis of the variation of
event-by-event fluctuations as control parameters are varied.

Once experimentalists vary a control parameter
which causes 
the freeze-out point in the $(T,\mu)$ plane to move toward, through, 
and then past the vicinity of the endpoint E, they 
should see all the signatures we describe first strengthen, 
reach a maximum,  
and then decrease, as a nonmonotonic function
of the control parameter.  It is important to have a control
parameter whose variation changes the $\mu$ at which
the system crosses the transition region and freezes out.
The collision energy is an obvious choice, since it is
known experimentally that varying the collision energy
has a large effect on $\mu$ at freeze-out.  Other possibilities
should also be explored.\footnote{If the system crosses the
transition region near E, but only freezes out at a much
lower temperature, the event-by-event fluctuations will not
reflect the thermodynamics near E.  In this case, one can
push freeze-out to earlier times and thus closer to E by
using smaller ions.\cite{SRS,laterfreezeout}} 

An example of nonmonotonic signatures in a different
but analogous context is
the rise and fall
in the number of large fragments
as a function of total observed multiplicity
in multifragmentation experiments~\cite{trautmann}
in low energy nuclear collisions. These experiments
allow us to confirm the existence and study the properties of another
critical point --- the end point of the first-order nuclear liquid-gas
transition (boiling of the nuclear matter liquid
to yield a gas of nucleons)~\cite{gilkes,trautmann}. 
This point occurs at a
temperature of order 10 MeV, much lower than the one we are
studying~\cite{SRS}.

The analogy which is perhaps most familiar is with the 
phenomenon of critical opalescence observed in most 
liquids, including water.
As the fluid cools down under conditions such
that it passes near the end point of the boiling transition, it
goes from transparent to opalescent to transparent as the
end point is approached and then passed.
This nonmonotonic phenomenon is due to the
scattering of light on critical long wavelength density fluctuations,
and thus signals the universal physics unique to the vicinity
of the critical point.  

The universal property of systems in the vicinity of a second
order critical point is the anomalous increase of thermodynamic
fluctuations of the order parameter and related observables.  
Here we consider a specific 
system, namely the hadronic matter created in a heavy ion collision
at the time interactions freeze out.   Our generic expectation
is that the {\it event-by-event} fluctuations of suitable
observables increase in the vicinity of a critical endpoint.  
In this paper we calculate 
the magnitude of the resulting nontrivial effects, and 
make predictions which, we hope, will allow experiments to 
find the endpoint E. 

It is clear that before we can achieve this goal we must develop
sufficient understanding of {\em non-critical} event-by-event
fluctuations. 
Large acceptance detectors, such as NA49 and WA98 at CERN,
have made it possible to measure important average
quantities in single heavy ion collision events.
For example, instead of analyzing the
distribution of charged particle
transverse momenta obtained by averaging over particles
from many events, we can now study the event-by-event
variation of the mean transverse momentum 
of the charged pions in a single event.
The event-by-event variation of
particle abundance ratios and even of the HBT radii are 
also becoming available.
Although much of this data still has preliminary status, with
more statistics and more detailed analysis yet to come, some general
features have already been
demonstrated. In particular, the event-by-event 
distributions of these observables are as perfect 
Gaussians as the data statistics allow, and the
fluctuations --- the width
of the Gaussians --- are small.\cite{trento}

This is very different from what one observes in $pp$ collisions, 
in which fluctuations are large.
These large non-Gaussian fluctuations clearly
reflect non-trivial quantum
fluctuations, all the way from the nucleon wave function to that of the
secondary hadrons, and are not yet sufficiently well understood.
As discussed in~\cite{GM92,GLR99}, thermal
equilibration in $AA$ collisions drives the variance of the event-by-event
fluctuations down, close to the value determined by the variance of 
the inclusive one-particle
distribution divided by the square root of the multiplicity.
In $pp$ physics one can hope to extract 
quantum mechanical information about the initial state
from event-by-event fluctuations of the final state;
in heavy ion collisions 
equilibration 
renders this an impossible goal. 
In $AA$ collisions, then, the new goal is
to use the much smaller, Gaussian
event-by-event fluctuations of the final state
to learn about thermodynamic properties at freeze-out. 

What can we learn from
the magnitude of these small fluctuations and their dependence on the
parameters of the collision?
Do they contain any more information than the corresponding
moments of one-particle inclusive
distributions?  Some of these questions have been addressed in 
\cite{Stodolsky,Shu_fluct} where it was pointed out that,   
for example, temperature fluctuations are related to heat capacity via
\be \label{cv}
{\langle (\Delta T)^2 \rangle\over T^2} = {1 \over C_V(T) },
\ee
and so can tell us about {\em thermodynamic} properties of
the matter at freeze-out. 
Similar ideas in \cite{Shu_fluct} 
relate fluctuations in the occupation
of certain momentum bins with $\partial \mu/\partial N$ and
the average quantum density in phase
space. Furthermore, Mr\'owczy\'nski has discussed the study of the
compressibility of hadronic matter at freeze-out via the
event-by-event fluctuations of the particle number \cite{Mrow1} 
and Ga\'zdzicki\cite{Gaz} and
Mr\'owczy\'nski\cite{Mrow2} have considered event-by-event
fluctuations of the kaon to pion ratio as measured
by NA49\cite{trento}.  

In this paper, we focus on observables constructed
from the multiplicity and the momenta of the charged particles in the final
state, as measured by NA49.  It should be possible
to use similar methods to analyze the
event-by-event fluctuation of other classes
of observables.  For example,
if it were possible to measure the baryon to pion ratio,
analyses analogous to those we discuss would lead
to the thermodynamic susceptibility 
$\partial^2 \Omega/\partial \mu^2$.  As the neutrons
are not observed, this analysis is not available.
However, event-by-event fluctuations of the kaon to pion ratio
may yield similar 
information. Another example is the data obtained by
WA98 on the
event-by-event fluctuation of the
charged particle to photon ratio.\cite{steinberg}
They find a Gaussian distribution, and therefore
constrain non-equilibrium processes in which
long wavelength disorientations of the chiral condensate
are excited, as these introduce non-Gaussianity.
We leave the extension of
the methods of this paper to the study of thermodynamic
implications of the NA49
Gaussian distribution of event-by-event $K/\pi$ ratios
and of the WA98 Gaussian distribution of event-by-event
$\pi^0/\pi^\pm$ ratios for future work.

Thermodynamic  relations like  (\ref{cv}) suggest the
following strategy.  Measure 
the mean transverse momentum 
of the charged pions in each event in an ensemble.
Since  the inclusive average of the transverse momentum of
pions from an ensemble of events reflects (although does
not equal) the temperature
of the ensemble, perhaps one can use $p_T$, the mean transverse
momentum of the pions in a single event,\footnote{We 
denote the mean transverse momentum 
of all the pions in a single event by $p_T$ rather than 
$\langle p_T\rangle$ because we choose to reserve $\langle \ldots \rangle$
for averaging over an ensemble of events.}
as a proxy for
the temperature of a single event, and so use (\ref{cv})
to obtain $C_V$.
One of the lessons of the results we present below is that this strategy
is too naive.  To see a sign of this, consider another
fundamental thermodynamic relation, namely that the event-by-event
fluctuations of the energy $E$ of a part of a finite 
system in thermal equilibrium are given by
\be
\label{energyfluctuation}
\langle (\Delta E)^2\rangle = T^2 C_V(T)\ .
\ee
For a system in equilibrium, the mean values of $T$ and $E$ 
are directly related by an equation of state $E(T)$;
their fluctuations, however,
have quite different behavior as a function of $C_V$, and
therefore behave differently when $C_V$ diverges at a critical point.
So, is the $C_V$-dependence of the event-by-event fluctuations
of $p_T$ like that of $\Delta T$ in (\ref{cv}) or
like that of $\Delta E$ in (\ref{energyfluctuation})? We will show
that $p_T$ fluctuations are not like either,
although their $C_V$-dependence is 
more similar to that of $\Delta E$, in the sense that
the fluctuations of $p_T$ grow at the critical point.

Most of our analysis is applied to the fluctuations
of the observables characterizing the multiplicity and 
momenta of the charged pions in the final state of a
heavy ion collision. 
There are several reasons why the pion observables
are most sensitive to the critical fluctuations. First, the
pions are the most numerous hadrons produced and observed 
in relativistic heavy ion collisions.
A second, very important reason, is that pions couple strongly to 
the fluctuations of the sigma field (the magnitude of
the chiral condensate) which is
the order parameter of the phase transition.  Indeed, the pions
are the quantized oscillations of the phase of the chiral
condensate and so it is not surprising that at the critical
end point, where the magnitude of the condensate is fluctuating
wildly, signatures are imprinted on the pions.  By Section 5,
we will have built up the technology needed to analyze
these signatures.

Before we outline the structure of the paper, the following
comment is in order. We assume throughout 
that freeze-out occurs from an equilibrated hadronic system.  
If freeze-out occurs ``to the left'' (lower $\mu$;
higher collision energy) of the critical end point E, it occurs
after the matter has traversed the crossover region
in the phase diagram. If it occurs
``to the right'' of E, it occurs after the matter has traversed
the first order phase transition.  This is the situation in 
which our assumption of freeze-out from an equilibrated system
is most open to question. First, one may imagine hadronization
directly from the mixed phase, without time for the hadrons
to rescatter.  Hadronic elastic scattering cross-sections are
large enough that this is unlikely.  Second, one may worry
that the matter is inhomogeneous after the first order
transition, and has not had time to re-equilibrate.
Fortunately, our assumption is testable.
If the matter were inhomogeneous at freeze-out, 
one can expect non-Gaussian fluctuations
in various observables \cite{heiselberg} which
would be seen in the same experiments that seek 
the signatures we describe.  We focus
on the Gaussian thermal fluctuations of an equilibrated
system, and study the nonmonotonic changes in these
fluctuations associated with moving the freeze-out point toward and then
past the critical point, for example from left to right as
the collision energy is reduced.

Although our central point is the analysis of the critical
fluctuations in the vicinity of the point E, 
we must first present an extensive
analysis of the noncritical fluctuations, which are
the background on top of which critical effects must 
be sought.  Thus, this paper is organized as follows:
Sections 2 and 3 analyze the background noncritical fluctuations;
Section 4 analyzes a particular (negative) contribution to 
the noncritical fluctuations
which disappears near the critical point; Sections 5 and 6
analyze the critical fluctuations themselves.

We begin in Section 2 by discussing the simplest case
we can imagine, namely the fluctuations in an ideal
Bose gas of pions.  This allows us to establish some 
notation and to explain several conceptual issues.
In particular, we explore the relation in this
simplest case between the ensemble (i.e.,
event-by-event) variance
and the variance of the inclusive one-particle distribution
obtained by averaging over particles from many events.
We also point out that the correlation between
the multiplicity and
an intensive observable, like the mean transverse momentum,
only receives contributions from nontrivial
effects such as  Bose enhancement, energy conservation
or interactions.  This correlation
is in general small, but we see in Sections 4 and
5 that it can increase by a large factor near
the critical point.
We derive results in Section 2 and throughout
which are valid in the thermodynamic limit. In an 
Appendix, we explain the subtleties of constructing
estimators for the relevant quantities using a finite
sample of events each with a finite number of pions.

Our goal in Section 3 is the inclusion of various
effects neglected in Section 2, except that
we continue to  assume that
freeze-out is {\it not} occurring in the vicinity of the
critical point.  We model the matter in a relativistic
heavy ion collision at freeze-out as
a resonance gas in thermal equilibrium, and begin by calculating
the variance of the event-by-event 
fluctuations of total multiplicity. 
Our result suggests that about 75\% of the 
fluctuations seen in the data
are thermodynamic in origin.
Our prediction is strongly dependent on the presence of the
resonances; had we not included them, our prediction would have been
significantly lower, farther below the data.

Fluctuations in extensive observables like the total multiplicity $N$
are sensitive to nonthermodynamic variation in the {\it initial} size
of the system which later thermalizes.  Sources of such variation
include: (i) the distribution of impact parameters; (ii) fluctuation
in the initial positions of the nucleons; 
(iii) quantum fluctuations
of the NN cross section~\cite{Baym_etal} described by the wave function of 
the nucleon, which can be thought of as fluctuations in the 
effective size of the nucleons at the initial moment of the
collision.  All these effects lead to fluctuation in the number of
spectator nucleons, and thus in the initial size of the interacting
system which later thermalizes.  We plan to evaluate the size of these
contributions to fluctuations in $N$ elsewhere.  In this paper, we
constrain the magnitude of these nonthermodynamic effects by comparing
thermodynamic predictions for the fluctuations in $N$ to the data.

We then turn to a 
calculation of the variance of the event-by-event
fluctuations of the mean transverse momentum, $p_T$.
This is an intensive variable and should, therefore, be
less sensitive to nonthermodynamic variations in the
initial size of the system.
We calculate numerically the thermodynamic contribution from ``direct pions'',
already present at freeze-out, and from the pions generated later
by resonance decay.  We include Bose effects and the effects
of flow and find both to be small. 
We compare our results to those found by NA49 for
central Pb-Pb collisions at $160$ AGeV, and find broad
agreement.  We do not attempt to include purely experimental
effects, such as those due to two-track resolution, and
so do not expect precise agreement.  Our goal is to
compare observed variance with thermodynamic expectations
and to see whether they are consistent.
Our results support the
general idea that the small fluctuations
observed in $AA$ collisions, relative to those in $pp$,
are consistent with the hypothesis that the matter in
the $AA$ collisions achieves approximate local thermal
equilibrium in the form of a resonance gas. 
Once data is available for other collision energies, centralities
or ion sizes, the present NA49 data and the calculations
of this section will provide an experimental and
a theoretical baseline for the study of variation as
a function of control parameters.

In Sections 4 and 5, we analyze how the proximity of
the critical endpoint to the freeze-out point is reflected
in the fluctuations.  We begin in Section 4 by making
the idealization that the pions which one observes
are an ideal Bose gas in thermal contact with a
heat bath which includes the sigma field. 
The heat capacity of this heat bath 
is therefore infinite at the critical point.  
This treatment neglects
the $\sigma \pi\pi$ coupling, which allows the critical
fluctuations of the sigma field to influence the pion
fluctuations directly, rather than just by thermal contact.  

The dominant effects of the critical 
fluctuations on the pions are the direct effects occurring
via the $\sigma \pi\pi$ coupling.  The idealization
of Section 4 is nevertheless useful, because it allows us to 
explain and illustrate an important point not made
clear in \cite{SRS} related to the practical application
of (\ref{cv}).  
The fluctuations of the temperature {\em depend} on what
``mechanical'' observable (such as the energy, for example) is 
measured, and how the measured observable is converted into 
a temperature.
In particular, these fluctuations depend on what part of a system
is used as a thermometer.
Eq.~(\ref{cv}) describes a particular case when
the whole system of interest 
is used as a thermometer. It requires us to use the
equation of state, $T(E)$, of the whole system of interest to translate the
energy, which is measured in this case, into the temperature~\cite{Landau}.
The fluctuations of ``mechanical'' variables, such as energy, {\em
increase} at the critical point, as in (\ref{energyfluctuation}).
Because $T(E)$ is singular at the critical point, 
the fluctuations of $T$ decrease, and vanish at the critical
point where $C_V\to\infty$.
It is a fact that what we measure
are the mechanical observables, and since we in general
only know $T(E)$ for simple systems we call thermometers,
we cannot apply (\ref{cv}) to the complicated system of interest.
We illustrate these points by evaluating the
fluctuations of several observables in 
an ideal gas of detected pions (the thermometer)
which is in thermal contact with an undetected non-ideal, possibly singular,
heat bath.  The effect we find vanishes at the critical
point, where the specific heat of the heat bath diverges
due to the fluctuations of the sigmas therein, 
and so provides a nonmonotonic signature.  
The effect involves a 
reduction in the fluctuations of the mean transverse
momentum of the pions. What makes it distinctive is that
it also involves an anti-correlation between fluctuations
of pion occupation numbers with different momenta.  We show that this 
phenomenon follows directly from energy conservation,
and conclude that it is much more robust than the
idealizations we use to describe it.
This signature is present when the system freezes out
far from the critical point, and is reduced near the critical point.
It should be observable in 
present data on central PbPb collisions at 160 AGeV,
even if freeze-out is not occurring near
the critical point in these collisions.

Section 5 describes what we believe to be the dominant
event-by-event signatures directly related to the divergent correlation
length which characterizes the critical point. 
We apply 
much of the technology built up over the preceding
sections in Section 5 to study the effect of the
interaction of the pions with the almost classical sigma
field.  We find a large increase in the fluctuations
of both the multiplicity and 
the mean transverse momentum of the pions. This increase
would be divergent in the infinite volume limit 
precisely at the
critical point. We apply finite size and finite time
scaling to estimate how close the system created in
a heavy ion collision can come to the critical singularity,
and consequently how large an effect can be seen
in the event-by-event fluctuations of the pions.  We conclude that
the nonmonotonic changes in the variance of the event-by-event
fluctuation of the pion multiplicity and momenta
which are induced
by the universal physics characterizing the critical point
can easily be between one and two orders of magnitude
greater than the statistical errors in the
present data.

Once we have analyzed the effects of the sigma field on 
the fluctuations of the pions, in Section 6 we ask what
becomes of the
sigmas themselves.  Assuming that freeze-out occurs
near the critical point, they are numerous at freeze-out 
and they can only decay later, 
once the sigma mass has risen above twice the pion mass.
This results in a nonmonotonic signature of the critical point
which can be observed even without an event-by-event analysis.
We calculate the momentum distribution 
of these low momentum pions produced in the delayed decays
of the sigmas. We close
by analyzing the enhanced event-by-event fluctuations of the 
multiplicity of these low momentum pions.

We end the paper with a summary of the different contributions
to the event-by-event fluctuations which we have analyzed,
and a more general look to the future. In striving to provide
analyses which will assist experimentalists to 
use the universal properties of the critical point
to learn its location, we hope that we have in addition provided 
a set of tools for event-by-event analyses of heavy ion collisions which
will prove useful in the study of the thermodynamics of QCD in
a variety of contexts in the future.

%% file: chap2.tex
\section{Thermodynamic Fluctuations in an Ideal Bose Gas }
\label{gas}

\subsection{The Basics}
\label{sec:basics}

We begin by recalling text-book facts about the thermodynamics
of an ideal Bose gas which are relevant to our event-by-event
analysis. Little in this section is new, but it is nevertheless a very
helpful exercise and will allow us to establish some notation.
The basic fact is that every quantum state of a system of
identical spinless Bose particles is completely characterized by a set of 
numbers, $n_p$ --- the occupation numbers for the one-particle 
states labeled by momenta $p$. All thermodynamic quantities are 
functions of these numbers and thus all we need to know is the 
fluctuations of $n_p$ from one member of the ensemble (one event) to 
another. The first step toward a characterization of
these fluctuations is the ensemble average of the
occupation number for the mode with momentum $p$, namely
\begin{equation}\label{barn}
\langle n_p \rangle = {1\over e^{\epsilon_p/T} - 1},
\end{equation}
where $\epsilon_p=\omega_p - \mu$ and, as usual, $\omega_p=\sqrt{p^2+m^2}$.
Next, we need the
deviation, $\Delta n_p = n_p - \langle n_p\rangle$, whose
mean square average in the ensemble is
given by
\begin{equation}\label{dn2}
\langle(\Delta n_p)^2 \rangle = T {\partial n_p\over\partial\mu}
= {e^{\epsilon_p/T}\over (e^{\epsilon_p/T} - 1)^2}
= \langle n_p\rangle (1 + \langle n_p\rangle) \equiv {v}^2_p.
\end{equation}
We have introduced notation ${v}^2_p$ for
this quantity which will be used frequently below.  
This expression is ``microscopic'', in the sense that it
is written for a single mode in momentum space. However,
it can be derived ``macroscopically'' as follows.
The fluctuations in the total particle number 
\be 
N=\sum_p n_p
\ee
are given by\cite{Landau}
\be
\langle (\Delta N)^2\rangle = T \left(\frac{\partial N}{\partial \mu}\right)_T
\ .
\ee
Because the fluctuations of different modes are statistically
independent, we can elevate this relation to the microscopic
form (\ref{dn2}), and indeed to
\begin{equation}\label{dndn}
\langle \Delta n_p \Delta n_k\rangle 
= \langle (\Delta n_p)^2 \rangle \delta_{pk}={v}^2_p\delta_{pk}.
\end{equation}
The correlator $\langle \Delta n_p \Delta n_k\rangle$ is 
the central quantity which we will calculate repeatedly
throughout this paper, as we proceed beyond the ideal
Bose gas.




The correlator in (\ref{dndn}) enters in the calculation of the
event-by-event mean square deviation of any generic thermodynamic
variable of the form
\be 
Q = \sum_p q_p n_p\ .
\ee
Indeed, since $\Delta Q \equiv Q - \langle Q\rangle 
= \sum_p q_p\Delta n_p$, we find that
\be\label{DeltaEfluctuation}
\langle (\Delta Q)^2\rangle = 
\sum_{pk} q_p q_k \langle \Delta n_p \Delta n_k\rangle =
\sum_p q_p^2 {v}^2_p.
\ee
The quantity $Q$ could be the total energy
\begin{equation}\label{EandDE}
E = \sum_p n_p \epsilon_p, 
\qquad \langle (\Delta E)^2\rangle = \sum_p \epsilon_p^2 {v}^2_p\ ;
\end{equation}
or it could be the total transverse momentum, $\sum_p (p_T)_p n_p$;
or simply the
total particle number
\begin{equation}
N = \sum_p n_p,
\quad \langle (\Delta N)^2\rangle = \sum_p {v}^2_p.
\end{equation}

For future reference we also give here an expression for the
heat capacity $C_V$ of the Bose gas at constant $V$ and $\mu$:
\begin{equation}
C_V = T \left(\partial S\over \partial T\right)_\mu = 
- T \left(\partial^2 \Omega \over \partial T^2\right)_\mu.
\end{equation}
Using the expression for the thermodynamic potential:
\begin{equation}
\Omega = T \sum_p \ln(1-e^{-\epsilon_p/T}).
\end{equation}
one finds
\begin{equation}\label{eqforCV}
C_V = {1\over T^2} \sum_p \epsilon_p^2 {v}^2_p.
\end{equation}
Comparing to (\ref{EandDE}) we find the well-known relation
\begin{equation}
\langle(\Delta E)^2\rangle = T^2 C_V\ ,
\end{equation}
which is valid for any system in equilibrium.

\subsection{Energy per Particle: Event-by-Event Average vs. Single-Particle 
Inclusive Average}

Let us now compute 
the fluctuation of an {\em intensive}
observable, such as the mean energy per particle $\epsilon = E/N$, where
$E$ and $N$ are extensive,
or $p_T$, the mean transverse momentum per particle in
a single event.  Analyzing the 
member-of-the-ensemble-by-member-of-the-ensemble fluctuations of
the mean energy per particle in a single member of the ensemble
is a good warmup.  We henceforth begin to refer to
members of the ensemble as events.
For small fluctuations (and 
$\Delta E/E\sim N^{-1/2} \ll 1$ is small) we can write
\begin{equation}
\Delta\left(E\over N\right) \approx  {E\over N}
\left({\Delta E\over E} - {\Delta N\over N}\right).
\end{equation}
Now, we square:
\begin{equation}\label{DE/Nsq}
\left(\Delta\left(E\over N\right)\right)^2 
= {1\over N^2} \left( (\Delta E)^2 
+ \left(E\over N\right)^2 (\Delta N)^2
- 2 \left(E\over N\right) \Delta E \Delta N \right).
\end{equation}
Then we average. We already know $\langle(\Delta E)^2\rangle$ and 
$\langle(\Delta N)^2\rangle$, but we also need
\begin{equation}
\langle \Delta E\Delta N\rangle = \sum_p \epsilon_p {v}^2_p,
\end{equation}
which is obtained in the same way as before.
Putting this all together, we find
\begin{equation}\label{de}
\langle(\Delta \epsilon)^2\rangle 
= {1\over \langle N\rangle^2} 
\sum_p (\epsilon_p - \langle\epsilon\rangle)^2 {v}^2_p.
\end{equation}

Let us now compare (\ref{de}) to the 
variance of the inclusive single
particle average energy per particle.  
To do this, we introduce the notation
\begin{equation}
\overline{q_p}^{\rm inc} \equiv \sum_p q_p \langle n_p\rangle/\sum_p
\langle n_p\rangle = \sum_p q_p \langle n_p\rangle/\langle N \rangle\ .
\label{incavgdefn}
\end{equation}
Whereas $\langle ... \rangle$ denotes an average over events
of some 
property of a single event, $\overline{ ^{\,}... _p }^{\rm inc}$ denotes
an inclusive average of a property of a single pion
over all pions in the ensemble of events, without reference
to in which event each pion occurs.  
It is more
convenient for theoretical purposes to work with occupation
numbers $n_p$, and the inclusive average is then done 
$n_p$ by $n_p$ as defined
in (\ref{incavgdefn}).  The subscript $_p$ on
the left hand side of (\ref{incavgdefn}) reminds us that the
average is done momentum bin by momentum bin:  it is $q_p$
which is being averaged, {\it not} $q=Q/N$.  However, the 
quantity $\overline{q_p}^{\rm inc}$ is $p$-independent.
Were we only interested in 
a quantity like $\langle \epsilon \rangle$,
there would be no need to take care with definitions
because averaging a single particle quantity pion
by pion is the same as first averaging it over an event,
and then averaging event-by-event:
\begin{equation}
\langle \epsilon \rangle = \langle E/N \rangle =
\overline{\epsilon_p}^{\rm inc}\ .
\label{ebevsincforeps}
\end{equation}
This is not true for fluctuations 
about the mean, as we see by using our definitions to
rewrite (\ref{de}) as
\begin{equation}\label{e=1p}
\langle(\Delta \epsilon)^2\rangle= {1\over \langle N \rangle}
\overline{(\epsilon_p - \langle \epsilon\rangle)^2 
(1+\langle n_p \rangle) }^{\rm inc}\ .
\end{equation}
The same formula holds if $\epsilon =E/N$ is 
replaced by any quantity of the form $q=Q/N$,
for example by the mean transverse momentum per event.

The lesson we learn from (\ref{e=1p}) is that up to the Bose 
enhancement factor $(1+\langle n_p\rangle)$,
the ensemble (alias event-by-event) fluctuations of {\em intensive}
quantities, such as the energy per particle, are indeed given by 
the variance of the single  
particle distribution $\overline{(\epsilon_p
- \langle \epsilon\rangle )^2}^{\rm inc}$ and the central limit
theorem which dictates the factor $1/ \langle N\rangle $. 
We see that the effect of the Bose factor is to increase
the variance of the event-by-event distribution relative 
to that of the inclusive distribution.

When we apply formulae like those we have just 
derived which are valid in the thermodynamic limit
to heavy ion collision data, we will need to 
construct estimators for the relevant  
quantities using a finite
ensemble of events, in which the number of particles in each 
event is also finite. We describe
how this should be done in an Appendix.

Having discussed the fluctuations of extensive and 
intensive quantities, we end this Section by considering
the cross correlation between an intensive observable and the extensive
observable $N$.
For example, let us calculate 
$\langle\Delta \epsilon \Delta N\rangle$.  Using ingredients we
have spelled out above, we find
\begin{equation}\label{inclexclcorr}
\langle \Delta \epsilon \Delta N\rangle =
\frac{1}{\langle N\rangle} \sum_p v_p^2 \left(\epsilon_p - 
\langle\epsilon\rangle\right)  = \frac{1}{\langle N\rangle} 
\sum_p \langle n_p\rangle^2 \left(\epsilon_p - 
\langle\epsilon\rangle\right)\ .
\end{equation}
Note that the terms proportional to $\langle n_p\rangle$ have cancelled,
and the remaining term, proportional to $\langle n_p\rangle^2$, 
is obviously due
to the Bose effect.  This result applies to any such correlation;
for example we could have used $p_T$ instead of $\epsilon$.
The lesson we learn is this: cross correlations
between $N$ and intensive observables are generally
small, because they receive no contribution if one takes
the classical ideal gas limit.   Recall that in (\ref{ebevsincforeps})
we find a dominant contribution to the event-by-event variation
coming from the variation of the inclusive single particle 
distribution. In (\ref{inclexclcorr}), this effect cancels 
and the remaining effects due to Bose enhancement dominate.
This means that the nontrivial effects on the pions 
due to their interactions and due to energy conservation
and thermal contact
with other degrees of freedom
only need to be larger
than the effects of Bose enhancement in order to
dominate this cross correlation.

%% file: chap3.tex
\section{Noncritical Thermodynamic Fluctuations\\ in Heavy Ion Collisions}

In this section we proceed to quantitative estimates of the
magnitude of noncritical event-by-event fluctuations in heavy ion
collisions. As an example of an {\em extensive} quantity we use
the total charged pion multiplicity of an event; as an example
of an {\em intensive} quantity we use the mean transverse
momentum $p_T$ of the charged pions in an event.
We compare some of our estimates to 
preliminary data 
from the NA49 experiment at CERN SPS on PbPb collisions
at 160 AGeV, and find broad agreement.
In this section, and throughout this paper,
we assume thermal equilibrium at freeze-out.  In this section, but not
throughout this paper, we assume that the system freezes
out far from the critical point in the phase diagram, and
can be approximated as an ideal resonance gas when it
freezes out. 
The results obtained
seem to support the hypotheses that most of the fluctuation
observed in the data is indeed thermodynamic in origin and
that PbPb collisions at 160 AGeV do not freeze out near the
critical point.

\subsection{Pion  Gas at Thermal Freeze-out and Bose Enhancement}
\label{sec:bose}

The observed spectrum of pions reflects the distribution of pion
momenta at the time of thermal freeze-out, namely the time at 
which the interaction rates fall behind the
expansion rate. After this time, one can approximately neglect 
energy/momentum exchange interactions and consider the momenta
of particles as frozen.   Freeze-out is by definition the time
at which the system ceases to be in thermal equilibrium.  
However, if the system has thermalized before it  freezes out,
then even after freeze-out one has a thermal distribution
of pion momenta,\footnote{As is very accurately 
the case for the cosmic microwave background radiation, 
ten billion years after its freeze-out.}
approximately with a single temperature
over the whole system.
This standard idealization at
this point seems sufficient to describe the data.  (Particles
which interact more weakly than pions freeze out earlier,
at a higher temperature. We leave such particles, together
with details of the dynamics of the freeze-out of the pions,
to future work.)

We start with the simplest model for 
the pions at freeze-out --- the ideal Bose gas. 
This allows us to
use the results of the previous section.  Later in this section,
we add pions produced by the decay of resonances
as well. 
The isospin degeneracy of the pions requires a small modification
to the formulae of the previous section.
Since only the momenta of
charged pions are observed,  we must only count $\pi^+$
and $\pi^-$. Because $\pi^+$ and $\pi^-$ are distinct,
the Bose enhancement factor is reduced
from $1+n_p$ to $1+(n_p/2)$, where $n_p$ counts the total number
of charged pions. This is the consequence of the fact that only 
identical pions can interfere.\footnote{It is 
easy to see that $\langle\Delta n^i_p\Delta n^j_k\rangle
= \langle n^i_p\rangle (1 + \langle n^i_p\rangle) \delta^{ij}\delta_{pk}$,
where $i,j=+,-$. On the other hand, from $n_p=\sum_i n^i_p$ it follows
that: $\langle n^i_p\rangle = \langle n_p\rangle/2$, and $\langle\Delta
n_p\Delta n_k\rangle = \langle n_p\rangle(1 + \langle n_p\rangle/2)
\delta_{pk}$.}

We begin by showing that the effect of the Bose enhancement 
is very sensitive to a nonzero pion chemical potential $\mu_\pi$
(not to be confused with the baryon number chemical potential $\mu$). 
Let us first remind the reader
why a nonzero $\mu_\pi$  may be needed.
The pion chemical potential is not a 
thermodynamic conjugate to any fundamentally
conserved quantity, and is the same for pions of all charges.
It is supposed to represent the over-population of pion phase space.
It allows for the possibility that even though the momenta
of the pions are in equilibrium at freeze-out, their number
density is not.
This arises because all reactions which can change the
number of particles, and thus keep
this quantity in equilibrium,  have small cross
sections at the relevant low energies. In contrast, 
elastic re-scattering is strongly enhanced by
resonances
(such as $\Delta,N^*$ for $\pi N$, $\sigma,\rho$ for $\pi\pi$ etc).
As a result, thermal equilibrium of momenta is maintained
to a lower freeze-out temperature, whereas chemical freeze-out
(below which particle numbers do not change) occurs somewhat
earlier.  There is therefore a window of time between
chemical and thermal freeze-out during which the system
evolves with fixed pion number; during this
time a pion chemical potential naturally develops.
At chemical freeze-out, $\mu_\pi=0$.  As the temperature
then 
continues to drop while the pion number remains fixed, $\mu_\pi$ increases.
%
%
For an overview of pion kinetics and references see \cite{kinetics}.
Practical calculations of the magnitude of the effect for heavy ion
collisions at CERN SPS can be found in \cite{HS}. The conclusion
inferred from this analysis is that the pions
in central PbPb collisions at SPS energies freeze out at a temperature
$T_{\rm f}\approx 120 {\rm ~MeV}$ with $\mu_\pi \approx 60$ MeV.

Now we return to the calculation of the Bose enhancement of fluctuations
of some generic single-particle intensive observable $q=Q/N$. 
If we use the notation $v_{\rm ebe}$ for the
event-by-event variance and $v_{\rm inc}$ for the variance of the
inclusive distribution:
\begin{equation}
v_{\rm ebe}^2(q) = \langle(\Delta q)^2\rangle \quad\mbox{ and }\quad
v_{\rm inc}^2(q) = \overline{(q_p - \langle q\rangle )^2 }^{\rm inc},
\end{equation}
and define the ratio 
\begin{equation}\label{Fdef}
F\equiv \frac{\langle N \rangle v_{\rm ebe}^2(q)}{v_{\rm inc}^2(q)}
\end{equation}
then we can write the result (\ref{e=1p}) for the ideal Bose 
gas as
\begin{equation}
F=F_B\equiv 
1 + \frac12 {\int d^3p (q_p - \langle q\rangle)^2 \langle n_p\rangle^2 
\over \int d^3p (q_p - \langle q\rangle)^2 \langle n_p\rangle}\ .  
\end{equation}
The factor of $1/2$ appears because, as discussed above, there
are two charged pions.  As we consider effects
not present in an ideal gas, we will find that the ratio $F$ is not
given simply by the Bose enhancement factor $F_B$. It is a 
product of $F_B$ and other factors which we estimate
later in this section and in subsequent sections.



The dependence of $F_B$
on $\mu_\pi$ is shown in Fig. 1 for $q=p_T=\sqrt{p_x^2+p_y^2}$.
(Note that $v^2_{\rm inc}$ does depend on the 
pion chemical potential as well.) 
We have also shown $F_B$ for
pions with $p_T<300$ MeV to demonstrate that restricting the
acceptance to low-energy pions results in a larger Bose enhancement
effect.
It is worth noting that for more central collisions the thermal
freeze-out temperature, $T_{\rm f}$, is lower because the 
system is larger and
freezes out later~\cite{laterfreezeout}. Therefore, $\mu_\pi$ should
be somewhat larger and the 
Bose enhancement effect should somewhat 
increase event-by-event fluctuations for more central collisions.
\begin{figure}[thb]
\label{fig_Fb} 
\centerline{\psfig{file=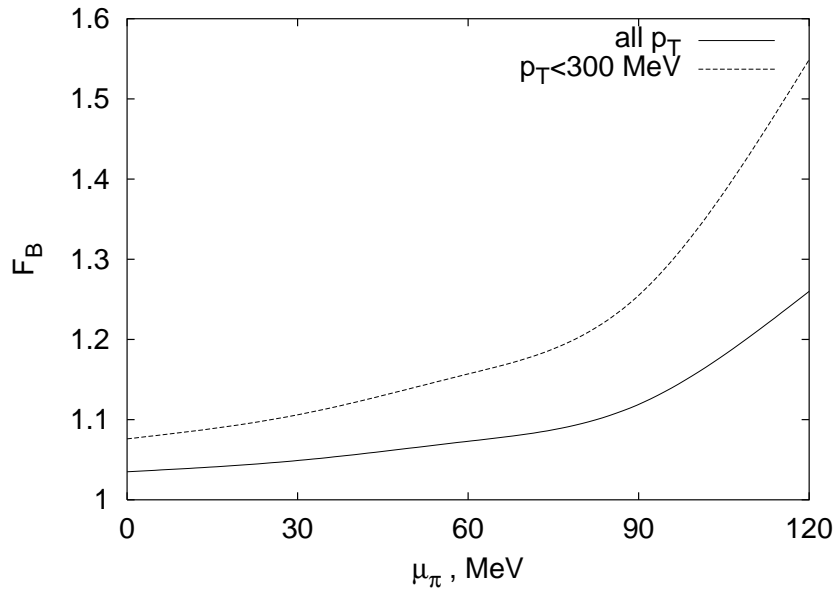,width=4.5in}}
\smallskip
\caption[]{Bose enhancement factor
$F_B=\langle N \rangle v^2_{\rm ebe}(p_T)/v^2_{\rm inc}(p_T)$
describing the contribution of the Bose effects to 
the fluctuations of the mean transverse momentum 
in an ideal Bose gas of pions.  $F_B$ is plotted 
as a function of the pion chemical
potential $\mu_\pi$, in MeV. The dashed
line shows the Bose enhancement factor if only pions
with low momentum $p_T<300$MeV are included.}
\label{fig:bose}
\end{figure}
To conclude, the Bose enhancement 
effect is sensitive to $\mu_\pi$,  and leads to an
increase in $v_{\rm ebe}$ by a factor of 
$\sqrt{F_B}$, which typically
results in an increase of the order of a few percent.

The effects of Bose enhancement on the variance of
the fluctuations of $p_T$ in an ideal Bose gas have been considered 
previously\cite{mrow3,trento}: 
our results are in quantitative agreement with theirs.
These authors
use the quantity 
\begin{equation}
\Phi_{p_T} = \langle N \rangle^{1/2} v_{\rm ebe}(p_T) 
- v_{\rm inc}(p_T)
\label{phiptdef}
\end{equation}
introduced in Ref. \cite{GM92} as a measure of the 
Bose enhancement effect.  As we discuss in the Appendix
and below, when $\langle N \rangle$ is finite one must
take care in defining $\Phi_{p_T}$ to use an appropriate
estimator for $v_{\rm ebe}(p_T)$.
To compare our results with theirs, note that
what we describe as
$\sqrt{F_B}=1.01$ corresponds to $\Phi_{p_T}= (0.01)v_{\rm inc}(p_T)$.
This already hints at what we will see below, namely that 
whereas $\Phi_{p_T}$ depends on the flow velocity through
$v_{\rm inc}(p_T)$, ratios like $F_B$ are much less sensitive to the effects 
of flow and are therefore more
easy to calculate.

\subsection{Contribution of Resonances}
\label{sec:res}

The hadronic matter produced in a heavy ion collision
 is {\em not} simply an ideal
gas of pions. 
A number of approaches to heavy ion collisions have
successfully treated the matter at freezeout as a
resonance gas in thermal equilibrium. 
Our analysis
of the fluctuations 
observed in present data lends support to this idea. 


Even the global properties of hadronic matter 
 are strongly affected by
resonances.  Although at temperatures of interest  
($T<T_c$) the 
Boltzmann factor $\exp(-M/T)$ for each resonance is small, 
it is partially compensated by the pre-exponential factor due to
the large number of states involved. 
One may recall here the
Hagedorn conjecture, that at a
certain temperature the contribution to the energy
density due to the resonances would diverge because of the
exponential growth of the density of states. Although
this does not happen in practice, because the chiral phase
transition occurs at a much lower temperature than any putative
Hagedorn transition,
one nevertheless finds \cite{Shu_72} that
when relevant resonances are included,
the energy density and pressure increase rapidly with $T$,
and can be fitted by a power law 
\be\label{resgas}
 \varepsilon(T) \sim P(T) \sim T^\kappa
\ee
with the power $\kappa\approx 6$ at zero baryon
density.\footnote{For nonzero
baryon density this effective power is even larger, but we will ignore
this since it is only important at much higher baryon
densities (and lower collision energies) than achieved
at the CERN SPS.} 
One also finds that the heat capacity, normalized 
to the number of pions (which means that we have
``decayed'' all resonances,  counting each $\rho$ meson as 2 pions, 
each $\omega$
as 3, etc.) is
\be
(C_V/N_\pi)_{\rm resonance\ gas}\approx 23,
\ee
at $T_f=120$ MeV, while for the ideal Bose gas of pions one has only:
\be 
(C_V/N_\pi)_{\rm pions}\approx 14 
\ee
at the same temperature.
This general observation \cite{Shu_fluct} already
suggests that the resonances may affect the fluctuations
considerably.


The resonances play another role in the problem. Those
which are present at freeze-out decay after freeze-out,
and by definition this means that the pions they produce
cannot rescatter. The pions observed in the data
are therefore a sum of (i) ``direct pions'' which were
pions at freeze-out and (ii) pions produced from
the decay of resonances after freeze-out.  Note that
the direct pions also originate from resonances,
in the sense that most of the low energy rescattering 
which occurs before freeze-out occur via resonances.
As we have discussed above, inelastic reactions which change
the number of participants freeze-out earlier than elastic
scattering. What this means is that the multiplicities of 
the pions and resonances, although thermal, should be fixed 
not
at the thermal freeze-out ($T_{\rm f}\sim 120$ MeV in PbPb) but 
earlier, at chemical freeze-out.  
Fits of SPS data on ratios of particle yields to
thermal models yield  
$T_{\rm ch}\sim 160-170$ MeV~\cite{PBM_etal}.\footnote{Note 
that this number is close to
the critical temperature obtained in lattice simulations
with zero baryon density\cite{latticeTc}.}

In the remainder of this subsection, 
we  investigate three effects of the resonances on the event-by-event
fluctuations of the extensive observable $N$, the number
of charged pions per event, and the intensive observable
$p_T$, the mean transverse momentum per event.  We first describe
all three effects briefly, and then describe the simulation
which we have used in order to investigate them.
The first (and largest) effect is a direct contribution
to the fluctuations of $N$, and indeed to any extensive
observable.  The multiplicity fluctuations in a 
classical ideal gas are characterized by 
$\langle (\Delta N)^2\rangle/\langle N \rangle=1$ and
for an ideal pion gas this ratio is 
$1+(1/2)\overline{\langle n_p\rangle}^{\rm inc}$ due to Bose
effects and is therefore a few percent larger than 1. 
This ratio is significantly larger for a resonance gas.
Each resonance decays into several pions
(for example,
$\rho\rightarrow 2\pi,
\omega,\eta\rightarrow 3\pi$, etc), 
and this means that when the resonances decay after freeze-out,
they significantly modify the statistics of pion number fluctuations.
If the resonances themselves are produced randomly\footnote{All resonances
are heavy enough that Bose enhancement for them can be neglected.},
with a Poisson multiplicity distribution, their decay products are not.
For example, if there were no direct pions and 
only one species of resonance
which always decayed into $d$ charged pions, the pions produced
in this ensemble would have 
$\langle(\Delta N)^2\rangle/\langle N\rangle=d$.

Resonances also affect the fluctuations of intensive observables,
like the energy per pion or mean transverse momentum $p_T$.
The second effect we analyze arises because pions produced
in resonance decays have a single-particle momentum
spectrum which is similar but not
identical to the thermal spectrum for the direct pions.
The products of resonance decay populate the low $p_T$ 
region of the spectrum somewhat more.\footnote{This effect is
qualitatively similar to the effect of a nonzero pion chemical
potential. A clear distinction between these effects in data
analysis is still lacking.}  
In order to estimate $v_{\rm ebe}(p_T)$, we must therefore include
the change in $v_{\rm inc}(p_T)$ introduced by the pions produced
by the decay of resonances after freeze-out.

The third effect of resonance decays is that they contribute additional 
kinematic correlations between their decay products, which then 
have no chance to rethermalize.  New terms arise in 
the correlator (\ref{dndn}) which describes the fluctuations
microscopically. 
For example, 
two-body decay (such as
$\rho\rightarrow \pi^+ \pi^-$ at rest) generates a term in 
the correlation function:
\be\label{resdecaycorrfn}
\langle\Delta n_p^+\Delta n_{k}^-\rangle 
= \delta_{p,-k} C(p),
\ee
where $C(p)$ is
proportional to the square of the pion fraction originating from
$\rho_0$ decays.  The result of all such terms will be
a change in $v_{\rm ebe}(p_T)$ which can be parameterized
as a new contribution $F_{\rm res}$ to the ratio $F$ of (\ref{Fdef}).
(That is, we now have $F=F_B F_{\rm res}.)$ Instead of attempting
to study all contributions like (\ref{resdecaycorrfn}) one
by one, we address this effect and the first two
by doing a  simulation.  We will see that the second and
third effects we have described are both small.

We have simulated a gas of pions, nucleons and resonances
in thermal equilibrium at freeze-out, including the 
$\pi$, $K$, $\eta$, $\rho$, $\omega$, $\eta'$, $N$, $\Delta$,
$\Lambda$, $\Sigma$ and $\Xi$,
and then simulated
the subsequent decay of the resonances. 
That is, we have generated an ensemble of pions 
in three steps:  (i) Thermal ratios of hadron multiplicities
were calculated  assuming equilibrium ratios at chemical
freeze-out. Following \cite{PBM_etal}, the values 
$T_{\rm ch}=170$ MeV and
$\mu_{\rm baryon}=200$ MeV  were used. (ii)  
Then, a program  generates 
hadrons with multiplicities determined at chemical
freezeout, but with thermal momenta as appropriate  
at the thermal freeze-out  temperature, 
which we take to be $T_{\rm f}=120$ MeV, with
$\mu_\pi=60$ MeV.  The last step (iii) is to
decay all the resonances, using the appropriate subroutine
from RQMD.\footnote{ We treat
particles which
decay by weak interactions as stable, which
raises an additional issue.
Experimentally,
some weak decays happen so quickly that they feed up into the
observed pion spectra. We treat these particles as
stable here;  we hope that 
experimentalists make 
the appropriate corrections to the data.} 
Under these conditions, more than half of  the observed pions
come from resonance
decays. 

We evaluate the variance of the fluctuations of the
multiplicity of the pions obtained from the resonance
gas as follows.  For each species in the 
resonance gas, we label the different decay modes
by an index $i$, and refer to the branching ratios for
the species $r$ as $b^i_r$.
For each decay we  
define $d^i_r$,
the number of charged pions produced. 
From the simulation, we obtain the multiplicity
of each resonance, $N_r$. The total multiplicity
of pions is $N_\pi=\sum_{r,i} d^i_r b^i_r N_r$ and the multiplicity
fluctuations are described by
\be\label{multflucprediction}
\frac{\langle (\Delta N_\pi)^2\rangle}{\langle N_\pi\rangle} = 
\frac{\sum_{r,i} (d^i_r)^2 b^i_r N_r}{\sum_r d^i_r b^i_r N_r} \approx 1.4 \ .
\ee
Bose enhancement increases this to 
$\langle (\Delta N_\pi)^2\rangle/\langle N_\pi\rangle\approx 1.5$.%
\footnote{
Event-by-event fluctuation in the resonance multiplicities $N_r$, as
may be computed, for example, in dynamical models in which the
resonances themselves are produced by decays of ``clusters'', may
result in a small further increase in this ratio.  
} 
We see that the resonances increase the multiplicity fluctuations by a
large factor, relative to the fluctuations of the direct pions alone.
We compare this result to what is seen in
NA49 data below.

\begin{table}
\begin{tabular}{|l||c|c|c|c|}\hline 
&no. of pions, & $\langle p_T \rangle$,& 
$v_{\rm inc}(p_T)$,&$v_{\rm inc}(p_T)/\langle p_T\rangle$\\
& $10^3$	&   MeV		&	MeV		& 		\\ \hline
``direct pions'' only 		& 541	& 283	& 189	& 0.67	\\ \hline     
pions from resonances only 	& 713    & 271	& 177 	& 0.65	\\ \hline  
all pions  			& 1254	& 276	& 183 	& 0.66	\\ \hline  
\end{tabular} 
\caption[]{Results of a numerical simulation of a resonance gas.
The results include the effects of the  
the correlations induced by resonance decays on the inclusive $p_T$
spectrum.  
The simulation itself does not include Bose enhancement effects,
and so can be thought of as the simulation
of a single event with $1.254\cdot 10^6$ pions with a mean $p_T$ of
276 MeV, or can be sliced up into smaller events.} 
%
\end{table}
We now turn to the resonance-induced contribution to the
fluctuations of the intensive observable $p_T$, which 
is much smaller. We begin with the effect on $v_{\rm inc}$.
Table 1 describes
the single-particle
inclusive distribution obtained from the simulation, assuming
uncorrelated particles in an equilibrium resonance gas at freeze-out.
We see that the resonances change $v_{\rm inc}(p_T)/\langle p_T\rangle$
only by a few percent.
The contributions of correlations induced by resonance
decay and of Bose enhancement to $F$ are not included.
The effects of flow are not included. We 
now discuss each in turn, and find that all yield small contributions
to $\langle N \rangle^{1/2} v_{\rm ebe}(p_T)/\langle p_T \rangle$
relative to $v_{\rm inc}(p_T)/\langle p_T\rangle$ 
which we have
evaluated in the table.

We have estimated $F_{\rm res}$ by slicing up the pions
from Table 1 into varying numbers (up to 2500) of events, 
and evaluating $F$.  Since Bose 
enhancement is not included in the simulation, the $F$ 
so obtained is just $F_{\rm res}$.  We find no statistically 
significant contribution to $F$, and conclude that $|F_{\rm res}-1|<0.01$.

We now use the results of Section 3.1 to incorporate 
Bose enhancement effects, after noting the connection
between Bose enhancement and resonance decay pions.  
There can be 
quantum interference between direct pions and resonance decay pions,
or among resonance decay pions. It is well known that 
all resonances can be approximately
separated into two groups: those which are short-lived 
and those which are long-lived.
The
former (e.g., $\rho$ and $\Delta$) have lifetimes much shorter than the
duration of pion radiation from the fireball (i.e., the
time over which freeze-out occurs). Therefore, their decay products
interfere with other pions. The decay products of
long-lived particles (e.g., $\omega$ and
$\eta$) can only
interfere with other pions if one selects pions with a very
small energy difference
$|\omega_1-\omega_2|<<\Gamma$, where $\Gamma$ is the 
width of the resonance. This is essentially impossible, and 
pions produced in the decay of 
long-lived particles therefore do not contribute to the Bose
enhancement factor. 
This means that $F_B-1$ should be multiplied by a factor
$(f_{\rm direct}+f_\rho+f_\Delta+...)^2$ where the $f$'s 
are the fractions of all
$\pi$ mesons coming from short-lived sources.  
This same fraction enters the HBT correlation function,
and is about 0.5.\cite{HeinzJacak}
So, we take $F_B-1 = 0.073$ for $\mu_\pi=60$ MeV from Fig.(\ref{fig:bose})
and reduce it by a factor of 0.5 yielding 
\begin{equation}
F_B=1.037\ , 
\end{equation}
and therefore conclude that the effect of Bose enhancement is a small
increase in 
$v_{\rm ebe}(p_T)$ by a factor of $\sqrt{F_B}=1.018$.

\subsection{The Effects of Radial Flow}
\label{sec:flow}

To this point, we have calculated the fluctuations
in $p_T$ as if the matter in a heavy ion
collision were at rest at freeze-out.  This is
not the case: by that stage  
the hadronic matter is undergoing a collective hydrodynamic
expansion in the transverse direction, and this
must be taken into account in order to compare
our results with the data.
A very important point here is that the fluctuations
in pion multiplicity  
are not affected by flow, and our prediction for them
is therefore unmodified.
Fluctuations in multiplicity ratios (e.g. $K/\pi$)
would also be unaffected.
However the event-by-event 
fluctuations of mean $p_T$ are certainly affected by flow.
The fluctuations we have calculated pertain to the rest
frame of the matter at freeze-out, and we must now boost
them. A detailed account of the resulting effects would
require a complicated analysis. Here we shall use 
the simple approximation\cite{SSH} that the effects of
flow on the pion momenta can be treated as a Doppler
blue shift of the spectrum: $n(p_T)\to
n(p_T\sqrt{1-\beta}/\sqrt{1+\beta})$.  This blue shift
increases $\langle p_T\rangle$, and increases $v_{\rm inc}(p_T)$,
but leaves the ratio $v_{\rm inc}(p_T)/\langle p_T\rangle$
(and therefore the ratio $v_{\rm ebe}(p_T)/\langle p_T\rangle$)
unaffected. 
This ratio (the fourth column in Table 1) 
is therefore a good quantity to compare to
experimental data, since our goal here is to extract information
about thermodynamics and not about flow.


However, event-by-event fluctuations in
the flow velocity $\beta$ 
must still be taken into account. 
This issue was 
discussed qualitatively already in \cite{Shu_fluct}, where it was
argued that this effect must be relatively weak. Here 
we provide the first rough estimate of its 
magnitude.
The magnitude of the flow velocity is proportional to the integral
of the pressure gradient over the expansion time. Thus the 
fluctuations of the flow velocity are determined by the pressure
fluctuations. The size of the pressure fluctuations is related
to the adiabatic compressibility by
the standard thermodynamic relation~\cite{Landau}
\begin{equation}
\langle(\Delta P)^2\rangle = -T\left(\partial P\over\partial V\right)_S.
\end{equation}
For the resonance gas equation of state this gives
\begin{equation}
{\langle(\Delta P)^2\rangle \over P^2}= {\kappa \over \kappa-1} {T \over P V}=
 {\kappa^2 \over \kappa-1} {1 \over S}.
\label{DeltaP}
\end{equation}
The entropy per pion in the ideal gas is around 2.4, and is larger
for the resonance gas. We shall take $S \approx 3 N_\pi $ for our
estimate.

Using the ``blue shift'' approximation we can write
\begin{equation}
p_T \approx p_T^{\rm rest} \sqrt{1+\beta\over1-\beta},
\end{equation}
where $p_T^{\rm rest}$ is the corresponding momentum in the rest frame
of the matter. The fluctuations of the observed momentum are then
related to the fluctuations in the rest frame (calculated above) and
the flow velocity fluctuations through
\begin{equation}\label{dpdbeta}
{\langle(\Delta p_T)^2\rangle\over p_T^2} = 
{\langle(\Delta p_T^{\rm rest})^2\rangle\over (p_T^{\rm rest})^2}
+ \langle(\Delta\beta)^2\rangle\ ,
\end{equation}
where we have neglected corrections which are
suppressed by $O(\beta)$ relative to $\langle(\Delta\beta)^2\rangle$.
The fluctuations in the flow velocity are given by
\begin{equation}\label{dbeta}
 \langle(\Delta\beta)^2\rangle = \beta^2 {\langle(\Delta P)^2\rangle \over
 P^2}\left( {\tau_{\rm micro}\over\tau_{\rm flow}}\right).
\end{equation}

The last factor on the right-hand side 
appears because the final velocity is
proportional to the time {\em integral} of the 
pressure gradient over the entire evolution
prior to freeze-out, and this integral is a sum over
uncorrelated fluctuations in time.
In a resonance gas one can discuss the
typical duration of a collision (the lifetime of a typical
resonance), and the time between collisions (the inverse of the
scattering rate). Both are close to the ``microscopic" time scale
$\tau_{\rm micro}\sim 1$ fm$/c$. The expansion duration relevant for
radial flow is actually much longer, $\tau_{\rm flow}\approx 10-20$
fm$/c$ for central Pb-Pb collisions. This means that for each
microscopic volume element one first does the time integral and obtains
a ``random walk'' factor $(\tau_{\rm
micro}/\tau_{\rm flow})^{1/2}\sim 1/4$ in $\Delta \beta$. 
Then, the
sum over uncorrelated
volume elements leads to a $1/\sqrt{V}$ or $1/\sqrt{N}$,
which we have already
seen in the expression (\ref{DeltaP}) for $\langle \Delta P \rangle$.

The flow velocity can be estimated for our purposes from the ratio
of $\langle p_T^{\rm rest}\rangle\approx 276$ MeV 
given in Table 1 and the experimental $\langle p_T\rangle\approx 376$ MeV
observed by NA49. Thus, $\beta\approx 0.3$. Finally, putting all the
estimates into eq.~(\ref{dbeta}) we find
\begin{equation}\label{betaflucts}
N \langle (\Delta\beta)^2\rangle \approx (0.1)^2.
\end{equation}
Note that although our estimate is uncertain at various points,
the result is very small.  Even if we have underestimated
the size of $\langle (\Delta \beta)^2 \rangle$ by a factor
of 4, the contribution to 
$v_{\rm inc}(p_T)/\langle p_T\rangle$   
would only be 0.02.  It is
quite clear that the great bulk of 
$v_{\rm inc}/\langle p_T\rangle$ is
thermodynamic, with the contributions of the fluctuations
in the flow velocity being negligible in comparison.

The largest uncertainty in our estimate for 
$v_{\rm inc}(p_T)/\langle p_T\rangle$ is not due to the fluctuations
in the flow velocity, which can clearly be neglected, but
is due to the velocity itself.  The blue shift approximation
which we have used applies quantitatively only to 
pions with momenta greater than their mass\cite{SSH}.
Because of the nonzero pion mass, boosting the pions
does not actually scale the momentum spectrum by 
a momentum independent factor. Furthermore, in a real
heavy ion collision there will be a position dependent
profile of velocities, rather than a single velocity $\beta$.
A more complete calculation of $v_{\rm inc}(p_T)/\langle p_T\rangle$
would require a better treatment of these effects in a hydrodynamic
model; we leave this for the future.

We obtain our final estimate of the magnitude of the event-by-event
fluctuations of the intensive quantity $p_T$ far from
the critical point as follows.
Using the estimate of $v_{\rm inc}(p_T)^{\rm rest}
/\langle p_T\rangle^{\rm rest}$ 
from Table 1 and Eqs.~(\ref{dpdbeta},\ref{betaflucts}), we estimate
that fluctuations in the flow velocity increase
$v_{\rm inc}(p_T)/\langle p_T\rangle$
from $0.66$ to $0.67$.  Multiplication by $\sqrt{F_B}$ then
yields
\be\label{finalprediction}
\frac{\langle N\rangle^{1/2} v_{\rm ebe}(p_T)}{\langle p_T\rangle} 
\approx 0.68\ ,
\ee
subject to the uncertainties introduced by the blue shift approximation.

\subsection{Comparison with NA49 Data and Outlook}

In this section we compare our results with the NA49 data from central
Pb-Pb collisions~\cite{trento} summarized in Table~\ref{tab:na49}.
\begin{table}[ht]
\begin{tabular}{|l||l|}\hline
number of events	&	98426			\\\hline
total number of charged particles	
&	26587685		\\\hline
\raisebox{-2pt}{$\langle N\rangle$}	&	270.13$\pm$0.07	\\\hline
$v_{\rm ebe}(N)$	&	23.29  $\pm$ 0.05	\\\hline
$\langle p_T \rangle$	&	376.75 $\pm$ 0.06 MeV	\\\hline
$v_{\rm inc}(p_T)$	&	282.16 $\pm$ 0.04 MeV	\\\hline
$v_{\rm ebe}(p_T)$			& 17.27 $\pm$ 0.03 MeV	\\\hline
\end{tabular}
\caption[]{Preliminary NA49 data\cite{trento}. The
charged particles are taken from the kinematic region
$0.005 < p_T < 2$ GeV and  $4 < y < 5.5$  (assuming $\pi$ mass). 
The events used are the $5\%$ most central of all events, with
centrality measured using a zero degree calorimeter.
The products from weak decays such as $\Lambda$'s and  $K^0$'s,
were only partially rejected with approximately 60\% rejection
efficiency. The errors are statistical only.}
\label{tab:na49}
\end{table}
As a first qualitative check of the predictions of our resonance gas
model, we can look at the multiplicity fluctuations.
It is clear that with no cut on centrality, one would see 
a very wide non-Gaussian
distribution of multiplicity determined by
the geometric probability of different impact parameters $b$.
Gaussian thermodynamic fluctuations can only be seen if
a tight enough cut in centrality is applied.  
The event-by-event $N$-distribution found by NA49 when
they use only the $5\%$ most central of all events,
with centrality measured using a zero degree calorimeter,
is Gaussian to within about $5\%$.  This cut corresponds
to keeping collisions with impact parameters $b<3.5$ fm.\cite{trento}
The non-Gaussianity
could be further reduced by tightening the centrality
cut further.  We now ask how well our resonance
gas describes the width of the (almost) Gaussian distribution.
From the data, we have
\begin{equation}
\frac{v^2_{\rm ebe}(N)}{\langle N \rangle } = 2.008\pm 0.009 \ ,
\end{equation}
which we should compare to our resonance gas prediction 
of 1.5.\footnote{In the NA49 data of Table 2, all 
charged particles
are counted whereas we have done our calculations assuming
that only the charged pions are observed. In our 
resonance gas model and in the data\cite{trento},
about 80\% of the charged particles in the final state are pions.
If we redo 
the calculation (\ref{multflucprediction}), but this time
define $d^i_r$ as the number of charged particles (pions,
kaons, protons) produced in the $i$'th decay of the $r$'th
resonance, we find that that 
$v^2_{\rm ebe}(N)/\langle N \rangle$ increases, but only 
by a few percent.  NA49 has demonstrated that it can study particle
identification event-by-event and it may therefore be possible
to analyze data on charged {\it pion} multiplicity in future.}
We therefore conclude that about $75\%$ of the observed
fluctuation is thermodynamic in origin.  The 
contamination introduced into the data by fluctuations in centrality 
could be reduced by
analyzing
data samples with more or less restrictive
cuts but the same $\langle N\rangle$, and extrapolating to a limit
in which the cut is extremely restrictive.
This could be done using
cuts centered at any centrality.  
In addition to fluctuations in centrality, there
is another experimental (as opposed to thermodynamic)
factor which could affect the agreement between
resonance gas predictions and the observed fluctuations.
The increase in the fluctuations
due to resonances can only be detected provided the detector acceptance 
is large enough
to ensure the detection of all (or most) of the decay products. NA49
seem to have coverage wide enough to satisfy this criterion and
a quantitative estimate of losses on its boundaries can easily be made.
Our resonance gas model predicts that
as the centrality cut is tightened, 
the ratio $v^2_{\rm ebe}(N)/\langle N \rangle$ should decrease toward 
a limit near 1.5.  

Although further work is certainly required, it is already apparent
that the bulk of the multiplicity fluctuations observed
in the data are thermodynamic in origin.  Because the multiplicity
fluctuations are sensitive to impact parameter fluctuations,
it may prove difficult to  explain  
their magnitude with greater precision
even in future. However, the fact that they are largely thermodynamic
in origin
suggests that the effects present near the critical point, which
we study in Sections 5 and 6, could result
in a
significant nonmonotonic enhancement of the multiplicity fluctuations.  
This would be of interest
whether or not the noncritical fluctuations on top of which
the nonmonotonic variation 
occurs are understood with precision.


Now, we proceed to $p_T$ fluctuations. As we explain in the Appendix,
in order to be sure that $F=1$ when there are no correlations
between pions, care must be taken in constructing an 
estimator for $v_{\rm ebe}(p_T)$ using a finite sample
of events, each of which has finite multiplicity. 
The appropriate 
prescription (\ref{vest}) is to
weight events in the event-by-event average by their
multiplicity.  This has not been done in Table 2. However,
we show in (\ref{estconv}) that we can use 
$\langle N \rangle$ and $v_{\rm ebe}(N)$
to change $v_{\rm ebe}(p_T)$ as required, and the result is
\begin{equation}\label{vebeconversion}
v_{\rm ebe}(p_T) = (17.27 \pm 0.03 {\rm MeV})\left(1 - \frac{1}{2} 
\frac{v_{\rm ebe}^2(N)}{\langle N\rangle^2}\right) 
= (17.21 \pm 0.03 {\rm MeV})\ .
\end{equation}
We use this henceforth.
We must  now compare  
\be
{\langle N\rangle^{1/2} v_{\rm ebe}(p_T)\over\langle p_T \rangle} 
= 0.751\pm 0.001\ 
\ee
to our prediction (\ref{finalprediction}) of 0.68.

We see that the major part of the observed fluctuation in $p_T$ 
is accounted for by the thermodynamic fluctuations
we have considered.  
A large part of the discrepancy is in
our prediction for the variance
of the inclusive single-particle 
distribution $v_{\rm inc}(p_T)$.
Our $v_{\rm inc}(p_T)/\langle p_T\rangle = 0.67$
is about $10\%$ lower than that in the 
data.\footnote{As we have already noted,
all charged particles are included in the data
whereas we have calculated the fluctuations
for the charged pions alone. We have checked that including
the protons and charged kaons from the resonance gas increases
our prediction for $v_{\rm inc}(p_T)/\langle p_T\rangle$
in the rest frame by only a few percent. 
This small increase in the ratio is likely further reduced
once the flow-induced increase in $\langle p_T\rangle$ 
for the kaons and protons is taken into account. 
Although it would be good to remove this uncertainty
completely by analyzing a data sample of pions alone, 
it is already clear that this is not the explanation
for the present 10\% discrepancy.}
First, this suggests
that there may be a small nonthermodynamic contribution
to the $p_T$-fluctuations, for example from fluctuations
in the impact parameter.%
\footnote{We expect that the fluctuations
of an intensive quantity like $p_T$ are less 
sensitive to impact parameter fluctuations than
are those of the multiplicity, and this seems 
to be borne out by the data.}
The other source of this discrepancy is the blue shift approximation.
We have applied a blue shift
factor such that $\langle p_T \rangle$ increases from $281$ MeV
in Table 1 to $377$ MeV as in the data, and in so doing
have obtained a value for $v_{\rm inc}(p_T)$ which is low by
$10\%$.  This may be a reasonable estimate for the
error which we have introduced by using the blue shift
approximation rather than a more sophisticated treatment
of the effects of flow on the spectrum, which we leave to
future work. Such a treatment is necessary before 
we can estimate how much of the 
$10\%$ discrepancy is introduced by 
the blue
shift approximation.  Future work on the experimental
side (varying the centrality cut) could lead to
an estimate of how much of the discrepancy is due
to impact parameter fluctuations.

We have gone as far as we will go in this paper 
in our quest to understand
the thermodynamic origins of the width of the inclusive
single particle distribution. 
Another very
important feature in the data is 
the value of the ratio 
of the scaled  event-by-event variation to the variance of the 
inclusive distribution:
\begin{equation}
\sqrt{F} = {\langle N\rangle^{1/2} v_{\rm ebe}(p_T)\over v_{\rm inc}(p_T)} 
= 1.002 \pm 0.002.
\label{dataratio}
\end{equation}
The difference between the scaled event-by-event variance
and the variance of the inclusive distribution is 
less than a percent.\footnote{As noted above, 
because $v_{\rm inc}(p_T)$ is scaled by the blue shift
introduced by the expansion velocity, so is 
$\Phi_{p_T}$. This makes $\Phi_{p_T}$ harder 
to predict than $F$.  However, for convenience, we note that if
one uses the experimental value of $v_{\rm inc}(p_T)$, a
value $\sqrt{F}=1.01$ corresponds to $\Phi_{p_T}=2.82$ MeV,
and the $\sqrt{F}$ in the data (\ref{dataratio}) corresponds
to $\Phi_{p_T}=0.6\pm 0.6$ MeV.}
This is a remarkable fact, since the contribution of the Bose
enhancement (see Section~\ref{sec:bose}) to this difference 
is almost an order of magnitude bigger ($\sqrt{F_B} -1$
is a few percent).
Therefore, there  must be  
some mechanism at work which compensates for the Bose
enhancement.
One possible mechanism is the effect of the two-track
resolution, diminishing the observed number of pions with very similar
momenta\cite{trento}.  This reduces the ratio
$F$, and
NA49 estimates that it is of comparable magnitude to the
Bose enhancement effect but with the opposite sign.  
We do not attempt to include either
this effect or the effect of final state Coulomb interactions
between charged pions
in our analysis, leaving that to the experimentalists.
However, we point out that in the next section we find
another possible origin of this effect.  We shall see that 
anticorrelations due to energy conservation and
thermal contact between the observed
pions and the rest of the system 
reduce
$F$, as long as the system does not freeze out near the critical
point.

In summary,  we have shown in this section that the 
assumption that the system is a thermal resonance
gas at freeze-out 
is in reasonable agreement with the 
magnitude of the observed event-by-event 
fluctuations in the pion
multiplicity and mean $p_T$.  We will see in Section 4
that the effects of energy conservation bring our
prediction for $\sqrt{F}$ into even better agreement with the data.
Of course, a number of issues we have touched upon
need further study: it cannot be otherwise for  the first
quantitative study of a new set of phenomena. 
The situation is, however, very encouraging.
First, RHIC detectors 
are very well-suited
to measurements of the fluctuations we have analyzed.
Second, although some of the interesting effects
are small ($F_B$, for example) with millions of 
recorded events all Gaussian widths can  be measured
to much better statistical accuracy than even the smallest
of the systematic effects we have discussed, and will
discuss in later sections.
Third, the interesting systematic effects can be studied
by varying the cuts made on the data.
For example, considering only
low-momentum pions one should find the effect of both 
the resonances and the Bose enhancement  (see Fig.(\ref{fig:bose}))
to be several times higher.  Also, $\mu_\pi$ and therefore
the Bose enhancement factor $F_B$ may be somewhat larger in central events.
Detailed study (varying centrality; varying cuts in $p_T$)
may allow experimentalists
to separate the effects of Bose enhancement from other effects we 
have described, and will describe later in this paper.
Fourth, one can significantly widen the types of 
fluctuations which are analyzed.  For example, one can
study new correlators like the event-by-event cross
correlation between multiplicity and mean $p_T$, $\langle \Delta p_T
\Delta N\rangle$. We saw in Section 2 that such cross correlation
results only from nontrivial effects.
Finally, it is important to note that we do not expect
any of the effects we have analyzed in this Section
to change significantly near the critical point.

Our analysis demonstrates
that the observed fluctuations
are broadly consistent with thermodynamic expectations, and 
therefore raises the possibility of large effects when
control parameters are changed in such a way that 
thermodynamic properties are changed significantly,
as at a critical point.  The smallness of the 
statistical errors in the data also highlights the possibility
that many of the interesting systematic effects we analyze
in this paper will be accessible to detailed study as
control parameters are varied.

%% file: chap4.tex
\section{Using an Ideal Gas of Pions as a Thermometer}
\label{sec:igasthermometer}

To this point, we have assumed
that the system does not freeze out close to the critical
point, and can be approximated at freeze-out as a
non-interacting ideal resonance gas.  In this Section,
we take a first step towards understanding
how the physics characteristic of the vicinity of the
critical point affects the event-by-event fluctuations.
Along the way, we quantify the effects of energy conservation
on the $p_T$-fluctuations. This leads to a small reduction
in $\sqrt{F}$ far from the critical point, which may
be required by the data (\ref{dataratio}).
In this Section, we
consider
only the ``direct pions'',
and as before we
treat them as an ideal Bose gas at freeze-out.
We further imagine that the pions are in thermal contact
with the ``rest of the system'', which is not directly observed
and which need not be ideal.
The rest of the system
includes the neutral pions, the resonances, the pions not
in the experimental acceptance and, most important, the 
order parameter or 
sigma field.  
If freeze-out occurs in the vicinity of the 
critical point, the thermodynamic properties of 
the sigma field (and therefore of ``the rest of the system'')
are singular.  In the analysis of this section, we
imagine that the observed pions are an ideal gas 
even for freeze-out in the vicinity of the critical point,
while the equation of state and susceptibilities of
the rest of the system become singular there.
Some of the universal critical indices characterizing this
singularity are discussed in \cite{SRS}.
The question we ask here is how the fluctuations of the 
pions are affected by being in thermal contact with the
rest of the system, particularly when the susceptibilities
characterizing the sigma field diverge.

A reader who is used to thinking about the $O(4)$ second
order transition may be concerned that we are treating
the pions and the sigma field so differently. The point
is that near the critical endpoint which we wish to analyze
(and which may occur in nature) the pions and sigma {\it are}
different.  The pions remain massive, while the sigma
mass vanishes and the long wavelength modes of the
sigma field undergo critical fluctuations and are almost classical.
The divergence of the specific heat of the system
as a whole is primarily due to 
the fluctuations of the sigma field.
The analysis of this section is therefore
a reasonable first step.  What it leaves out, of course,
is the fact that the pions, although not massless,
do interact strongly with the sigma field and
are therefore not an ideal gas.  
We are neglecting 
the direct effects
of the pion-sigma coupling.  Once these are included,
it is not possible to make a clean separation between
ideal pions and singular rest of the system.  We analyze
the consequences of the pion-sigma coupling in Section 5.


\subsection{Thermometers, Temperature Fluctuations and Heat Capacity}

Let us step back and recall the text-book formalism describing
the measurement of temperature. 
A thermometer should be
a simple system which has been already
calibrated, in the sense that we can relate
its total energy $E$ to its temperature $T$ via a function
$T(E)$ which we already know.  Instead of $E$, 
we could also use any other
mechanical observable, like for example the volume of the liquid in a liquid
thermometer.
An ideal pion gas makes a very good thermometer because it is a simple
system with a known equation of state. Having the equation of state,
we read off $T$ by measuring a mechanical observable, such
as $E$. 

If the mechanical observable fluctuates, so will the
measured temperature. In particular, if we measure the total energy,
which fluctuates as $\langle(\Delta E)^2\rangle =T^2C_V$ in the canonical
ensemble the temperature $T(E)$ will also fluctuate, with
$\langle(\Delta T)^2\rangle$ given by (\ref{cv}). 
Note that given that we measure a mechanical
observable $E$ rather than $T$, in order to find the result (\ref{cv})
we must know $T(E)$.  This is possible for a thermometer like
an ideal pion gas, but may not be possible for the system one 
wishes to study using the thermometer.
One of
the questions we address in this section is when we measure the energy of the
thermometer only (instead of measuring the energy of 
the whole system) which $C_V$ is
relevant: that of the thermometer, that of the rest of the
system, or a combination.

Suppose now we use our thermometer $B$ to measure the temperature of
another system $A$. The measurement consists of bringing the two
systems in thermal contact. If the resulting system $A+B$ is closed, 
thermal equilibrium will result. By ergodicity, the 
thermodynamic ensemble will consist of all the states with the 
same energy, $E_{A+B}$, taken with equal probability weight.
Although the total energy does not fluctuate, the
energies of the subsystems $E_A$ and $E_B$ do, subject to
a constraint $E_A+E_B=E_{A+B}$. The probability that
the subsystem $B$ has energy $E_B$ is proportional to
the number of states, $\Gamma$, of the system $B$ with energy $E_B$ times
the number of states of $A$ with energy $E_A=E_{A+B}-E_B$:
\begin{equation}\label{gammas}
\Gamma_{A+B}(E_{A+B}) = \sum_{E_B} \Gamma_A(E_{A+B}-E_B) \Gamma_B(E_B). 
\end{equation}
Both $\Gamma_{A,B}$ are exponentially growing functions of their
arguments (and also the size of the system) and their product on the
right-hand side of (\ref{gammas}) has a sharp maximum at some value of
$E_B$. Introducing the entropy $S$ as $S(E) = \ln\Gamma(E)$,
we can write for the value of $E_B$ at the maximum:
\begin{equation}\label{TA=TB}
0 = {d\over dE_B} \left[S_A(E_{A+B}-E_B) + S_B(E_B)\right] = 
- {dS_A\over dE_A} + {dS_B\over dE_B} = -{1\over T_A} + {1\over T_B},
\end{equation}
since the temperature is, by definition, $1/T = dS/dE$.
We recover the text-book result that the temperatures of two
systems in equilibrium are equal. Measuring $E_B$, and using
the known function $T_B(E_B)$, we find this (common) temperature.

Of course, it is not necessary that the system $A+B$ is rigorously closed.
In practice it is sufficient that the
rate of the thermal equilibration between $A$ and $B$ is faster 
than the rate of thermal equilibration of $A+B$ with the environment.

So far we have only discussed the mean value of $E_B$ and, consequently,
the mean temperature. The size of the fluctuations of $E_B$ is
given by the width of the maximum in $\Gamma_A(E_{A+B}-E_B) \Gamma_B(E_B)$. 
We need the second derivative:
\begin{equation}\label{secondderivative}
{d^2\over dE_B^2} \left[S_A(E_{A+B}-E_B) + S_B(E_B)\right] =
{d^2S_A\over dE_A^2} + {d^2S_B\over dE_B^2} = -{1\over T^2} \left(
{1\over C_A} + {1\over C_B}\right),
\end{equation}
where $C_{A,B}$ are the heat capacities of the systems $A$ and $B$.
Thus we find for $\Delta E = \Delta E_B = -\Delta E_A$:
\begin{equation}\label{dE}
\langle(\Delta E)^2\rangle = T^2 \left(
{1\over C_A} + {1\over C_B}\right)^{-1}.
\end{equation}
The importance of the result (\ref{dE}) is that the thermometer $B$ allows
us not only to measure the temperature of the system $A$, 
but also the heat capacity
of the system $A$!\footnote{One example of such a measurement  in a
simple lattice system can be found in \cite{Creutz}.} 
In order to make such a measurement, we must watch the
fluctuations of $E_B$ in addition to $\langle E_B\rangle$.

Another consequence of (\ref{dE}) is that when $C_A \gg C_B$ we
recover the result for the canonical ensemble (\ref{cv}). What is
important is that the heat capacity $C_V$ appearing in (\ref{cv}) in
this case is that of the thermometer itself, $C_B$, and not that of the
measured system, $C_A$.

Now, suppose that the system $A$ has a thermodynamic singularity at
some temperature, as a result of which $C_A\to\infty$.  
This is
precisely the situation which arises near the critical point
in the idealization of this section: the ideal pion thermometer $B$
is in thermal contact with a system $A$ with divergent
susceptibilities.
Equation
(\ref{dE}) tells us that the fluctuations of the energy, which are
equal in $A$ and $B$ due to the conservation 
of energy in $A+B$, will
{\em increase} as we approach the critical point where $C_A$
diverges. What happens to the temperature fluctuations?  Remember,
that we do not measure the temperature directly, but use the equation
of state $T(E)$ to read it off from the value of $E$.  If we used the
equation of state of the system $A$, $T_A(E_A)$, the fluctuations of
$T_A$ would {\em decrease} and vanish at the critical point as discussed
in~\cite{SRS}, because $C_A=dE_A/dT_A=\infty$.  However, the equation
of state of the system $A$ is not known to us. Indeed, we are trying
to learn about it doing our measurements. 
If we instead use the equation of
state of the thermometer $T_B(E_B)$ which 
is nonsingular, the fluctuations of $T_B$ will
increase because the fluctuations of $E$ do, 
and will approach the value determined by (\ref{cv}) with
$C_V=C_B$.  

Note that the temperatures of the systems $A$ and
$B$ determined through their respective equations of state are
different on the event-by-event basis. This is not in contradiction
with thermodynamics which only requires the {\em mean} values
to agree as in (\ref{TA=TB}).

Returning to our idealized system at freeze-out we want to
use pions we observe as a thermometer, $B$. The rest of the system,
which includes all the other particles, including pions not ending up
in our detectors, we consider as system $A$. 
The singularity of the heat capacity occurs in $C_A$, while $C_B$
is the heat capacity of the ideal gas and is regular. 
Nevertheless this singularity affects the fluctuations of the
pions through (\ref{dE}).  The effect of the singularity
in $C_A$ is an increase in the fluctuation $\langle(\Delta E)^2\rangle$.
If one were able to use $T_A(E)$ to define $T$, one would
find that fluctuations in $T_A$ would decrease at the
critical point.  Using $T_B(E)$, or any practical definition
of a temperature, leads to fluctuations in $T$ which, like those
in $E$, increase.
Since what we measure is always a mechanical thermodynamic observable,
like the total energy $E$, 
or the energy per particle, or the transverse
momentum per particle, etc., it is not in fact necessary
to do a translation to the temperature variable to detect
a singularity. It is easier
to look directly at the fluctuations of observable quantities.
To this we now turn.

\subsection{The Microscopic Correlator}
\label{sec:corrCACB}

As discussed in Section~\ref{sec:basics}, the mean square variations of
thermodynamic observables in the pion gas are determined by the
microscopic correlator: $\langle \Delta n_p \Delta n_k\rangle$. 
Once we find
this correlator we can then use it to
calculate any fluctuations of interest.

For the case of the canonical
ensemble this correlator is given by (\ref{dndn}) which leads to:
\begin{equation}
\langle (\Delta E)^2\rangle = \sum_p \epsilon_p^2 {v}^2_p = T^2 C_B,
\end{equation}
as in (\ref{EandDE}) and (\ref{eqforCV}). This corresponds to the case
$C_A=\infty$, where $B$ is a (grand) canonical ensemble. In the
case when $C_A$ is finite (\ref{dE}) tells us that the correlator
$\langle \Delta n_p \Delta n_k\rangle$ should change.
A simple derivation of this correlator given below 
yields the result
\begin{equation}\label{dndncorr}
\langle \Delta n_p \Delta n_k\rangle = {v}^2_p \delta_{pk}
- 
{{v}^2_p \epsilon_p {v}^2_k \epsilon_k \over 
T^2C_A + \sum_p {v}^2_p \epsilon_p^2}\ .
\end{equation}
This result is easy to understand intuitively and it passes many
nontrivial checks. When $C_A\gg C_B = \sum_p \epsilon_p^2 {v}^2_p/T^2$ 
the second term in
(\ref{dndncorr}) is negligible and we recover (\ref{dndn}). On the
other hand, when $C_A=0$, the system $B$ is closed and the total
energy $E=\sum_p \epsilon_p n_p$ cannot fluctuate. Accordingly,
$\sum_p \epsilon_p \langle \Delta n_p \Delta n_k\rangle =0$ in this
case. Note that the correlation is negative as it should be, since
finiteness of $C_A$ suppresses fluctuations of $E$, which means that if
one $n_p$ increases, others are more likely to decrease.  This negative
correlation is therefore a direct consequence of energy conservation,
and should persist even in systems which are less ideal than
the one we are analyzing in this Section.

The microscopic correlator (\ref{dndncorr}) determines the
fluctuations of many observables.  For example, by convolving
it with $\epsilon_p\epsilon_k$ as in (\ref{DeltaEfluctuation})
one can
derive the result (\ref{dE}). This is yet another check
of (\ref{dndncorr}).  Note also
that the correlation term in (\ref{dndncorr}) is down by a factor of
$1/V$ (since $C_{A,B}\sim V$), where $V$ is the volume of the system.  This
is also easy to understand: the restriction on some linear combination
of $n_p$'s imposed by energy conservation 
affects each individual $n_p$ little if the number of
$n_p$'s (i.e., the size of the system) is large.  However, the
contribution of this term to fluctuations of extensive or cumulative
quantities is not small, as equation (\ref{dE}) shows. This is
due to the absence of the Kronecker delta in the second term.


We now turn to the derivation of the result (\ref{dndncorr}).
The uncorrelated fluctuations given by formula (\ref{dndn}) follow
from the factorizable probability distribution%
\footnote{A careful reader may note that eq.~(\ref{dn2}) literally
implies that $\Delta n_p \sim V^0$, 
as far as the thermodynamic limit
$1/V$ power counting is concerned. $n_p$ is also 
of order $V^0$.  If it were the case that $\Delta n_p \sim n_p/\sqrt{V}$,
our assumption that the fluctuations of the occupation
numbers $n_p$ are Gaussian
would be immediately justified.
Instead, the fluctuations of the
occupation numbers $n_p$ are not necessarily Gaussian. This can be
cured by considering, in place of $n_p$, the sum of
occupation numbers of a set of modes in a cell $(\Delta p)^3$
centered at $p$ in momentum space, where $\Delta p$ is fixed as
$V\to\infty$. Since the number of modes in such a set is
$(\Delta p)^3V={\cal O}(V)$ and the modes fluctuate independently, the
central limit theorem will apply and make fluctuations of such
``smeared'' $n_p$ Gaussian. Practically, we always convolve
$n_p$ with a smooth function of $p$.
Instead of displaying the smearing of $n_p$ explicitly in our
notation, we can instead just treat the fluctuations of $n_p$ 
as if they are Gaussian, because this will not
affect any of the quantities
calculated by convolving $n_p$ with a smooth function of $p$.}
\begin{equation}
dP(n_p) = \prod_p dn_p \exp\left\{-{1\over2{v}^2_p}(\Delta n_p)^2\right\}.
\end{equation}
The energy $E=\sum_p \epsilon_p n_p$ in such an ensemble fluctuates
according to $\langle(\Delta E)^2\rangle = \sum_p {v}^2_p \epsilon_p^2 =
T^2 C_B$. 

Now, if we bring this system into thermal contact with the
system A, according to (\ref{gammas},\ref{TA=TB},\ref{secondderivative}) 
the probability receives an additional factor: 
$\exp[-(\Delta E)^2/(2T^2C_A)]$. For example, if $C_A=0$ it
becomes a delta-function, meaning that the system B is closed itself,
and the energy cannot fluctuate. So, we write:
\begin{eqnarray}
dP(n_p) &=& \left(\prod_p dn_p\right) 
\exp\left\{-\sum_p{1\over2 v_p^2}(\Delta n_p)^2\right\}
\exp\left\{-{1\over2T^2C_A}\left(\sum_p \epsilon_p \Delta n_p\right)^2\right\}
\nonumber\\ &=&
\left(\prod_p dn_p\right)
\int d\lambda \exp\left\{-\sum_p{1\over v_p^2}(\Delta n_p)^2
+\lambda\sum_p \epsilon_p \Delta n_p\right\}
\exp\left\{T^2C_A{\lambda^2\over2}\right\},
\end{eqnarray}
where we have introduced a Lagrange multiplier $\lambda$. The integration
over $\lambda$ should be done along the imaginary axis for convergence.

Completing the squares we find:
\begin{eqnarray}
dP(n_p,\lambda) &=& d\lambda \left(\prod_p dn_p\right)  \exp\left\{
-\sum_p{1\over2{v}_p^2}(\Delta n_p - \lambda{v}_p^2\epsilon_p)^2
\right\}
\nonumber\\ &&\times
\exp\left\{\left(T^2C_A+\sum_p{v}_p^2\epsilon_p^2\right){\lambda^2\over2}\right\}.
\end{eqnarray}
Now we see that:
\begin{equation}
\left\langle\left(\Delta n_p - \lambda{v}_p^2\epsilon_p\right)
\left(\Delta n_k - \lambda{v}_k^2\epsilon_k\right)\right\rangle 
= {v}_p^2\delta_{pk},
\end{equation}
\begin{equation}
\langle\lambda^2\rangle 
= - \left(T^2C_A+\sum_p{v}_p^2\epsilon_p^2\right)^{-1} = 
{-1\over T^2 (C_A+C_B)},
\end{equation}
and
\begin{equation}
\left\langle\lambda\left(\Delta n_k - \lambda{v}_k^2\epsilon_k\right)
\right\rangle=0 \ ,
\end{equation}
{}from which we find:
\begin{eqnarray}
\langle\Delta n_p\Delta n_k\rangle &=& {v}_p^2\delta_{pk}
+ \langle\lambda\Delta n_p\rangle {v}_k^2 \epsilon_k 
+ \langle\lambda\Delta n_k\rangle {v}_p^2 \epsilon_p
- \langle\lambda^2\rangle {v}_p^2 \epsilon_p{v}_k^2 \epsilon_k\nonumber\\ 
&=&
{v}_p^2\delta_{pk} 
+ \langle\lambda^2\rangle {v}_p^2 \epsilon_p{v}_k^2 \epsilon_k
\nonumber\\ &=&
{v}_p^2\delta_{pk} - {1\over T^2}
{{v}_p^2 \epsilon_p{v}_k^2 \epsilon_k\over C_A + C_B}.
\end{eqnarray}
Q.E.D.

Now, armed with the equation (\ref{dndncorr}), we can calculate
all other fluctuations in our ideal Bose gas $B$ in contact with
the system $A$.

\subsection{Application: Fluctuations of Mean $p_T$}
\label{sec:thermalcontactflucts}

As an example of the application of the formula for the microscopic
correlator (\ref{dndncorr}) we analyze 
the fluctuations of an intensive variable
in the ideal Bose gas of pions which we denote $q$,
$q=Q/N$.  We wish to see how the fluctuations of $q$
are influenced by the fact that the pions are in thermal
contact with a system with (possibly singular) heat capacity 
$C_A$.
We shall be interested in a particular
case where $q$ is the mean transverse momentum $p_T$, but shall 
use the more general notation both for later convenience and
to make contact with Section 2 in which we discussed $\epsilon=E/N$,
another possible $q$.

Starting from an equation similar to (\ref{DE/Nsq}), averaging, and 
using the correlator (\ref{dndncorr}) instead of (\ref{dndn}) we obtain:
\begin{eqnarray}
\langle(\Delta q)^2\rangle 
&=& \left\langle\left(\Delta\left(Q\over N\right)\right)^2 \right\rangle
\nonumber\\
&=& {1\over \langle N\rangle^2} \left\{
\sum_p {v}^2_p (q_p-\langle q\rangle )^2 
- {1\over T^2 ( C_A + C_B) }
\left[\sum_p{v}^2_p \epsilon_p (q_p-\langle q\rangle)\right]^2
\right\}\ .\label{qcacb}
\end{eqnarray}
The first term on the right-hand side is the same as in (\ref{e=1p}) 
with $q=\epsilon_p$. This is the main contribution
to $\langle(\Delta q)^2\rangle$. We have seen that
these thermodynamic fluctuations can be described
using the variance of the inclusive
single-particle distribution and the Bose enhancement factor. 
The second,
{\em negative}, term in (\ref{qcacb}) 
is the effect of the anti-correlation (second term in
(\ref{dndncorr})) induced by energy conservation
and thermal contact with
the system $A$. This term would be nonzero even if $C_A$ were zero. 
In this case, it would describe the effects of energy
conservation on the fluctuations of $q$ in the system
$B$. Thermal contact with the system $A$ reduces
this term, but it remains important as long as $C_A$ is
comparable to $C_B$.  It vanishes
at the point where $C_A$ diverges.

In a heavy ion collision, the heat capacity of the pion
gas $C_B$ is a sizable fraction of the total heat capacity $C_A+C_B$. 
The effect (\ref{qcacb})
can therefore decrease the fluctuations
of $p_T$, countering the Bose enhancement. This effect
will be reduced as we approach the critical point where $C_A$
diverges. This will lead to an increase in the event-by-event 
fluctuations of $p_T$ as compared to 
the variance of the inclusive single-particle spectrum.

To make the comparison with the Bose enhancement effect
estimated in Section~\ref{sec:bose} 
easier, we shall express the strength of the
effect of the thermal contact in terms of the ratio, $F_{T}$ 
of the whole expression
in curly brackets in (\ref{qcacb}) to the first term in this expression.
For the fluctuation of mean transverse momentum per event, i.e., for
$q=p_T$ we find 
\begin{equation}
F_T \approx 1 - {0.12\over C_A/C_B + 1} \ .
\label{FTresult}
\end{equation}
for $T=120$ MeV and $\mu_\pi =0$.  
We see that the effects of energy conservation and thermal
contact on the fluctuations of an intensive quantity
like $p_T$ are smaller than the effects on the fluctuations
of the energy in (\ref{dE}).  Several obstacles make it difficult
to use (\ref{FTresult}) quantitatively.  First, some dilution
of the effect is to be expected because less than half of
the pions which are observed are direct.
Second,
it is a little bit difficult to know how to estimate 
$C_A/C_B$, because we have analyzed such an idealized
situation. 
The system $A$ should 
certainly include the neutral pions in the same region of rapidity
as the observed charged pions; however, should the pions at different
rapidities be included? The total number of pions, neutral
and charged, in a central event at the SPS at 160 AGeV is about
ten times larger than the number of charged pions per
event used in NA49's present analysis.   This suggests
that $C_A/C_B$ is at the very most 10.
If we take $C_A/C_B\sim3$ for orientation, which can be 
justified if one assumes that 
$C_A$ includes the heat capacity of the resonances and
that of the neutral pions in the same region of phase
space as the observed pions, 
we find $F_T-1$ of the order
of $-3\%$, before taking into account the dilution by non-direct
pions. 
The effect is comparable in magnitude to the Bose enhancement,
acts in the opposite direction, and should be 
reduced near the critical point at which $C_A$ diverges.

A divergent specific heat is only possible in an infinite system.
In Section 5 we will estimate
that in a realistic heavy ion collision, finite size effects suggest that 
near the critical point the
sigma contribution to $C_A$ 
could be as much as a factor of $\sim 6^2$ larger than the contribution
of a typical light degree of freedom. This suggests that $C_A$ could
easily increase by as much as an order of magnitude at the critical point, 
reducing the anti-correlation in $\langle \Delta n_p \Delta n_k \rangle$
and the negative contribution to $F_T$ by the same factor.

The effects of thermal contact can be distinguished from other
effects, like those of finite two-track resolution,
which also counter the Bose enhancement effect because of the
specific form of the microscopic correlator (\ref{dndncorr}).
The effect of energy conservation and 
thermal contact introduces an {\em off-diagonal} (in $p$
$k$ space) anti-correlation.  Although our estimate of
the magnitude of the effect suffers from a variety of
uncertainties introduced by the idealizations
used throughout this Section, the existence of
this off-diagonal anti-correlation is robust.
It arises simply due to energy conservation:
when one $n_p$ fluctuates up others must fluctuate
downward, and it is therefore more likely that $n_k$ fluctuates
downward.   If $C_A$ is increased, then the system A can more 
easily supply the energy needed for the upward fluctuation
in $n_p$, and the anti-correlation between $n_p$ and $n_k$ is reduced.  
Preliminary analysis by NA49 suggests that
some amount of such anti-correlation is observed in the data\cite{Trainor}.
It will be interesting to compare the magnitude 
of any effect observed in
the data with our estimates. We leave this to future work.
If it is possible to separate this effect from 
other effects because it is an off-diagonal anti-correlation,
then a measurement of this effect would yield an estimate
for the effective value of the ratio $C_A/C_B$ at freeze-out.

Note that $F_T$ increases near the critical point, but it
increases towards a finite value (namely 1.) 
In contrast, in Section 5 we will explore effects 
which result in the {\em divergence} of an analogously
defined ratio $F_\sigma$ at the critical point.

\subsection{Another Application: $\tilde T$}
\label{sec:ttilde}

In this subsection we introduce another measure of the temperature of
the pion gas, $\tilde T$. 
Our new
variable $\tilde T$ is well-defined on a single event,
and has the property that $\langle \tilde T \rangle$ is related to the 
slope parameter.  We have found that although
$\langle p_T \rangle$ is related to $T$,
the fluctuations $\langle (\Delta p_T)^2\rangle$ are
not at all like the fluctuations $\langle (\Delta T)^2\rangle$
in (\ref{cv}).
We now show that $\langle (\Delta \tilde T)^2\rangle$
also does not behave quite like $\langle (\Delta T)^2\rangle$.
The reader should expect this, since we argued on general
grounds that (\ref{cv}) can only be obtained from a
mechanical observable if the equation of state of the
system $A$ is known.  Still, it is nice to confirm
this using an example of an observable which is
a less straightforward intensive quantity than just
$\epsilon=E/N$ or $p_T$.


We define for each member of
the ensemble (i.e., for each event) {\em independently}:
\begin{equation}\label{chiT}
\chi^2(T) = \frac12 \sum_p (n_p - n_p^0(T))^2 {1\over {\sigma}_p^2},
\end{equation}
where 
\begin{equation}
n_p^0(T) = {1\over e^{\epsilon_p/T} - 1},
\end{equation}
and ${\sigma}^2_p$ is some function of $p$ which
we can choose for convenience later.
Then for each event individually we can define a temperature, $\tilde
T$, which is found by minimizing $\chi^2(T)$ for this event:
\begin{equation}
\left[d \chi^2(T) \over dT\right]_{T=\tilde T} = 0.
\end{equation}
It is clear that mean value of $\tilde T$ over all events for the
ideal Bose gas will coincide with the actual temperature of the gas $T$.
But, since $\tilde T$ is defined 
for a single event, rather then for the whole
ensemble, it fluctuates!

As before, the fluctuations in $\tilde T$ are determined by
fluctuations of $n_p$. For small fluctuation $\Delta
\tilde T$ we can write:
\begin{eqnarray}\label{dndt1}
0 &=& \Delta \left[d \chi^2(T) \over dT\right]_{T=\tilde T} 
=  \Delta \left[\sum_p (n_p^0(T) - n_p) {\partial n_p^0\over\partial T}
{1\over {\sigma}_p^2}\right]
\nonumber\\ &&
= \sum_p \left(\Delta T {\partial n_p^0\over\partial T} - \Delta n_p\right)
{\partial n_p^0\over\partial T} {1\over {\sigma}_p^2},
\end{eqnarray}
where we omit the tilde on $T$.
Note that even if ${\sigma}^2_p$ contains dependence on $T$, it will be
always multiplied by $(n_p-n_p^0)$ which is zero to the relevant order
in the size of the fluctuation.
Now we need:
\begin{equation}\label{dn/dt}
{\partial n_p^0(T)\over\partial T} = {1\over T^2} \epsilon_p
n_p^0(1+n_p^0) = {1\over T^2} \epsilon_p{v}^2_p,
\end{equation}
according to (\ref{dn2}).
We rewrite (\ref{dndt1}) as:
\begin{equation}\label{dndt2}
{\Delta T\over T^2} \sum_p \epsilon_p^2 {v}^4_p {1\over {\sigma}_p^2}
= \sum_p \Delta n_p \epsilon_p {v}^2_p {1\over {\sigma}_p^2}.
\end{equation}

We can carry on with an arbitrary ${\sigma}^2_p$, but let us
 make the following choice: ${\sigma}_p^2={v}^2_p$.
This choice makes a lot of sense if one recalls that in the standard definition
of $\chi^2$ one divides each square deviation term by its
normal square deviation (which is usually obtained from experimental error,
and which here we know to be ${v}^2_p$ from the fluctuations of
$n_p$). This choice simplifies formulas.

Now we square (\ref{dndt2}), average over events, and restore the tilde
on $T$:
\begin{equation}
{\langle(\Delta \tilde T)^2\rangle\over \tilde T^4} 
\left[\sum_p \epsilon_p^2 {v}^2_p\right]^2
=  \sum_p \sum_k \epsilon_p \epsilon_k \langle\Delta n_p\Delta n_k\rangle,
\end{equation}
which, according to (\ref{dndncorr}) means:
\begin{equation}
{\langle(\Delta \tilde T)^2\rangle\over \tilde T^2} 
= {1\over C_B} {C_A\over C_A + C_B}. 
\end{equation}
We see that fluctuations of $\tilde T$,
like those of $p_T$, increase towards
the critical point of the system $A$, where $C_A\to\infty$,
approaching a finite constant. When $C_A$ is infinite, the system $B$
(the Bose gas) is in the canonical ensemble, and the fluctuations
of $\tilde T$ are given precisely 
by (\ref{cv}), with the specific heat
$C_B$ in the denominator.\footnote{The 
heat capacity $C_V$ in this case is that at fixed $\mu$.
It is remarkable that if we define $\tilde T$ by simultaneously
fitting {\em two} variables in (\ref{chiT}), $\tilde T$ and $\tilde \mu$,
the resulting $\tilde T$ will again fluctuate according to (\ref{cv}),
but with heat capacity at constant $N$. We leave this as an
instructive exercise for our reader.}


Different definitions of the temperature $\tilde T$ can be devised (using
different choices of ${\sigma}^2_p$). They will lead to different 
temperatures for a given event, which are the same in the mean 
(and equal to $T$), but different
in the size of their fluctuations. 
All these fluctuations will
increase somewhat at the critical point, but will not
diverge there as they are controlled there by $C_B$ (or
some other property of the thermometer $B$) which is nonsingular.



\subsection{Two Further Applications: $\langle (\Delta N)^2 \rangle$
and $\langle \Delta N \Delta p_T\rangle$}

Once we understand how some physical effect
influences the microscopic correlator 
$\langle \Delta n_p \Delta n_k \rangle$, we can calculate
the fluctuations of many different observables. The task
then is to look for observables in which the effect of interest
is large, and which are of practically utility in the sense
that they are easily accessible to experimental analysis.
We give two further simple examples here.

In Section 4.3, we analyzed the fluctuations of an intensive
quantity, $p_T$, and obtained the expression (\ref{qcacb}) for
$\langle (\Delta p_T)^2\rangle$.  Similarly, we can use 
the microscopic correlator (\ref{dndncorr}) to analyze the
fluctuations of the extensive quantity $N$, and obtain
\begin{equation}\label{dNdNthermal}
\langle (\Delta N)^2\rangle=
\sum_p {v}^2_p 
- {1\over T^2 ( C_A + C_B) }
\left[\sum_p{v}^2_p \epsilon_p \right]^2 \ .
\end{equation}
The first term is the ideal Bose gas result, and the second
term is the correction due to thermal contact and energy conservation.
For $T=120$ MeV, $\mu_\pi=0$, the effect of the second term is
to multiply $\langle (\Delta N)^2\rangle$ by a factor
of $[1-0.20/(1+C_A/C_B)]$.  Note, however, that the multiplicity
fluctuations of the pions obtain from the resonance gas
which we analyzed in Section 3 are dominated by the 
pions from those resonances which decay into more than
one pion.  Doubling the contribution of the direct pions
to $\langle (\Delta N)^2\rangle$ in the calculation (\ref{multflucprediction})
only increases $\langle (\Delta N)^2\rangle$ by $10\%$.  
The effect of thermal contact and 
energy conservation on the direct pions seen in (\ref{dNdNthermal})
is therefore a very small contribution to the total 
$\langle (\Delta N)^2\rangle$ of (\ref{multflucprediction}).

We saw at the end of  Section 2  that cross correlations between
intensive observables and $N$ are of interest, because
they vanish in a classical ideal gas.  We therefore use
the microscopic correlator (\ref{dndncorr}) to calculate
\begin{equation}
\langle \Delta N \Delta q \rangle =
{1\over \langle N\rangle} \left\{
\sum_p {n}^2_p (q_p-\langle q\rangle )
- {1\over T^2 ( C_A + C_B) }
\left[\sum_k{v}^2_k \epsilon_k\right] 
\left[\sum_p{v}^2_p \epsilon_p (q_p-\langle q\rangle)\right]
\right\}.
\end{equation}
For $q=p_T$, $T=120$ MeV, $\mu_\pi=0$ we find 
\begin{equation}
\frac{\langle \Delta N \Delta p_T \rangle}{\langle p_T\rangle}
= -0.021\left(1+{10.\over C_A/C_B + 1}\right)\ .\label{dNdpTthermal}
\end{equation}
As we saw in Section 2, correlations like this arise only 
due to nontrivial effects, and are
generally small. In this case, we see
that (for $C_A\sim 3 C_B$)
the effect of energy conservation 
and thermal contact is $\sim 2.5$ times as
large as
that due to Bose enhancement.  This suggests that this
correlation would be a very interesting quantity to use to look for 
the critical point.  It is small in magnitude, but even after the dilution
of the direct pions by those produced in resonance decays 
are taken into account, it
may change by
a large factor 
near the critical point where $C_A\rightarrow 0$.

In conclusion, the effects of thermal contact and energy
conservation on the pions could either be found
directly, by detecting the anti-correlation in the microscopic
correlator $\langle \Delta n_p \Delta n_k \rangle$.  Or, the resulting effects on
$\langle (\Delta p_T)^2 \rangle$, $\langle (\Delta \tilde T)^2 \rangle$,
$\langle (\Delta N)^2 \rangle$ or $\langle \Delta p_T \Delta N \rangle$
which we have estimated
may be discovered, likely by seeing them change as control parameters 
are varied.

%% file: chap5.tex
\section{Pions Near the Critical Point: Interaction with the Sigma Field}

In the previous section, we made the assumption that the ``direct
pions'' at freeze-out could be described as an ideal Bose gas.
We do not expect this to be a good approximation if
the freeze-out point is near the critical point.
The sigma field is the order parameter for the
transition and near the critical point it therefore develops large critical
long wavelength fluctuations. 
These fluctuations are responsible for singularities
in thermodynamic quantities.  In the previous section,
we analyzed this situation by pretending that the only
effect on the pions was due to thermal contact with
a heat bath with divergent susceptibilities.  In this
Section we take the next logical step, and consider
the effect of the classical critical fluctuations on the pions
through the $\sigma\pi\pi$ coupling.  
It would be strange
if, as in the previous
section, the properties of the pions remained regular in
the thermodynamic limit in the presence of the nonanalytic
behavior of the sigma field.  We will see that the fluctuations
of both the multiplicity and 
the mean transverse momentum of the pions do in fact diverge
at the critical point.

We then estimate the size of the effects in a heavy
ion collision. This requires first estimating the strength
of the coupling constant $G$, and then taking into account
the finite size of the system and the finite time during
which the long wavelength fluctuations
can develop.
The pion fluctuations induced by the $G\sigma\pi\pi$ interaction
are divergent and are therefore 
the dominant fluctuations in an infinite system.
In the finite system of interest, we find that the
momentum fluctuations  are large enough
to be easily detectable, 
but not so large as to seriously jeopardize
our treatment, which considers the 
effects of the interaction only to lowest order.
It is for this reason that 
we have first analyzed all effects other
than those introduced by the $G\sigma\pi\pi$ interaction,
and now add these effects in.  The multiplicity fluctuations
are large enough that in this case a treatment which goes beyond
lowest order in $G$ seems called for. We leave this
to the future.


\subsection{Microscopic Correlator}
\label{sec:corrG}

As before, we shall derive the expression for the microscopic ``master''
correlator, $\langle{\Delta n_p\Delta n_k}\rangle$, which can then be used 
to calculate fluctuations of various observables. 
We neglect the
effects considered in the previous Section, as they can
be added to the effects of this Section at the end.
We concentrate on the fluctuations of
the sigma field, the fluctuations of 
the pion occupation numbers, and 
the $\sigma\pi\pi$ coupling. The long wavelength fluctuations
of the sigma field which are responsible for the singular effects
of interest are classical.


The effective potential $\Omega$ determines the probability distribution
of the classical field $\sigma$ through
\begin{equation}\label{sigmaprobdist}
dP(\sigma) = d\sigma \exp\left\{ - \frac{\Omega(\sigma)}{T}\right\}.
\end{equation}
This equation can actually be thought 
of as the definition of $\Omega(\sigma)$.
The effective potential is extensive, but for convenience we
set the volume $V=1$ in the calculations to
follow, although we will restore it 
explicitly in our results.
(Note that throughout previous sections, we had set
$V=1$ implicitly.  The momentum sum 
$\sum_p$ should always be read as $V\int d^3p/(2\pi)^3$.)
Let us consider small
fluctuations of the field $\sigma$ around the minimum of $\Omega(\sigma)$.
We can then expand the effective potential $\Omega(\sigma)$
around $\sigma=0$. The first terms will be 
\begin{equation}\label{omegaphi}
\Omega(\sigma) = {m_\sigma^2\sigma^2\over2} + G\sigma:\pi^2: 
+ {\cal O}(\sigma^3)\ ,
\end{equation}
where we have temporarily omitted terms independent of $\sigma$ (such as
$m_\pi^2\pi^2/2$).%
\footnote{Clearly, the fluctuations of $\sigma$ are
not small. We shall proceed with the assumption that 
the higher-order terms in $\Omega(\sigma)$ yield
subleading contributions to the singular effect we seek. 
We shall
return to this point in Section~\ref{sec:finsiz}. Also note
that we consider only the zero momentum mode of the field
$\sigma$. This can be justified in a diagrammatic approach, which can
also handle nonzero momentum modes of $\sigma$.}
The second term is the interaction between sigmas and pions. 
The coupling $G$ 
has the dimensions of mass, and
its magnitude near the critical
point will be estimated below. The notation $::$
signifies tadpole subtraction: $:\pi^2: =\pi^2 - \langle{\pi^2}\rangle$,
which makes sure that the minimum of $\sigma$ is not shifted as
we shall see below.
(The notation $\pi^2$ itself is itself somewhat symbolic, as it 
represents $\int d^3x \pi(x)\pi(x)$.)
Thus we have:
\begin{equation}
dP(\sigma) = d\sigma \exp\left\{ - {m_\sigma^2\sigma^2\over2T} 
- {G\over T}\sigma:\pi^2: \right\}
\end{equation}

Now, the field $\pi$ also fluctuates. Let us determine the
corresponding (joint) probability
distribution. In the previous section we used the probability
distribution for the occupation numbers, and we begin
by translating the fluctuations of the field $\pi$ into 
fluctuations of the occupation numbers.
We write, doing the usual Fourier
transform:
\begin{equation}
\pi^2 = \sum_p |\pi_p|^2.
\end{equation}
We can relate the Fourier components $\pi_p$ to the
occupation numbers $n_p$. It is clear that $n_p\sim |\pi_p|^2$.
The coefficients can be determined, for example, by using
\begin{equation}
Z = \int {\cal D}\pi \exp\left\{
-\int_0^{1/T}dt\int dV \left[ 
\frac12(\partial_\mu \pi)^2 + \frac12m_\pi^2 \pi^2 \right]\right\} =
Z_{T=0} \prod_p \left(1-e^{-\omega_p/T}\right).
\end{equation}
Differentiating $\ln Z$ with respect to $m_\pi^2$ we find
\begin{equation}
\langle{\pi^2}\rangle = \sum_p {1\over \omega_p} \langle{n_p}\rangle,
\end{equation}
up to the temperature independent vacuum contribution (equal to 
$\sum_p\omega_p/2$) from $\ln Z_{T=0}$.
So we have
\begin{equation}
\pi^2 = \sum_p {1\over \omega_p} n_p.
\end{equation}
Note now that $\langle{\pi^2}\rangle = \sum_p \langle {n_p}\rangle/\omega_p
\ne0$. So, unless we subtract $\langle{\pi^2}\rangle$ the minimum of $\sigma$
will be shifted from the origin (this subtraction will also take care of
the vacuum fluctuations). We have:
\begin{equation}
\pi^2 -\langle{\pi^2}\rangle = \sum_p {\Delta n_p\over \omega_p}.
\end{equation}
Now, putting everything together, we find the joint probability
distribution for the sigma field and for the pion occupation numbers:
\begin{equation}
dP(\sigma,n_p) = d\sigma \left(\prod_p dn_p\right) \exp \left\{
-\sum_p {1\over 2{v}^2_p} (\Delta n_p)^2 
- {G\sigma\over T} \sum_p {\Delta n_p\over \omega_p}
- {m_\sigma^2\over 2T}\sigma^2 \right\}.
\end{equation}
This is a very important formula which will allow us to calculate
the fluctuations. 

The measure $dP(\sigma,n_p)$ is Gaussian, which is very helpful.
Completing the squares, we find
\begin{eqnarray}\label{phinp}
dP(\sigma,n_p) &=& d\sigma \left(\prod_p dn_p\right) \exp\left\{
-\sum_p {1\over 2{v}^2_p} 
\left(\Delta n_p + {G\sigma\over T}{{v}^2_p\over\omega_p}\right)^2
\right\}
\nonumber\\ && \times \exp \left\{
- \left({m_\sigma^2\over 2T} - {G^2\over T^2}\sum_p 
{{v}^2_p\over 2\omega_p^2}\right)\sigma^2
\right\}
\end{eqnarray}
Before we make the final and the simplest step, let us make two side
notes.

The equation (\ref{phinp}) shows that the interaction with $\sigma$
shifts mean occupation numbers by
\begin{equation}\label{npshift}
\delta \langle{n_p}\rangle = - {G\sigma\over T}{{v}^2_p\over\omega_p}\ .
\end{equation}
As the reader might have guessed already, this must be due to the
shift of the mass of the pions linear in $\sigma$, which can be seen
from (\ref{omegaphi}):
\begin{equation}\label{mpishift}
{\delta m_\pi^2 \over 2} = G\sigma. 
\end{equation}
It is trivial to evaluate the change in $\langle{n_p}\rangle$
induced by the change of the mass. Since $\epsilon_p = \sqrt{p^2
+m_\pi^2} -\mu$) we find:
\begin{equation}
\delta \langle{n_p}\rangle = -
\langle{n_p}\rangle(1+\langle{n_p}\rangle) 
{\delta\epsilon_p\over T} 
= - {v}^2_p {1\over T} {\delta m_\pi^2\over 2\omega_p},
\end{equation}
which is the same as (\ref{npshift}) with (\ref{mpishift}).
The fluctuations of the sigma field will have further
affects on the pion occupation numbers, but these are
higher order in $G$ and we neglect them here.

The second side note is more important. We see from (\ref{phinp})
that the mass of the $\sigma$ field is corrected by the fluctuations
of the pions:
\begin{equation}
\tilde m^2_\sigma = m^2_\sigma - 
{G^2\over T^2}\sum_p {{v}^2_p\over \omega_p^2}.
\end{equation}
Diagrammatically, this corresponds to the thermal one-loop diagram
$\sigma\to\pi\pi\to\sigma$. The physical mass
of the sigma is $\tilde m_\sigma$, to the order
in which we are working. This is the mass which vanishes
at the critical point. We shall omit the tilde in the following.

Finally, we can read off the following expectation values 
from the probability distribution (\ref{phinp}):
\begin{equation}
\left\langle{\left(\Delta n_p 
+ {G\sigma\over T}{{v}^2_p\over \omega_p}\right)
\left(\Delta n_k 
+ {G\sigma\over T}{{v}^2_k\over \omega_k}
\right)}\right\rangle = {v}^2_p\delta_{pk};
\end{equation}
\begin{equation}
\langle{\sigma^2}\rangle = {T\over m_\sigma^2};
\end{equation}
\begin{equation}
\langle{\sigma\Delta n_p}\rangle = 
- \langle{\sigma^2}\rangle {G\over T}{{v}^2_p\over \omega_p}.
\end{equation}
This gives
\begin{equation}\label{dndnsigma}
\langle{\Delta n_p\Delta n_k}\rangle = {v}^2_p\delta_{pk}
+ {1\over m_\sigma^2}{G^2\over T}
{{v}^2_p{v}^2_k\over \omega_p \omega_k}.
\end{equation}
We see that the coupling of the pions to the sigma field leads to a {\em
singular} contribution to the correlator of the pion fluctuations
as we approach the critical point at which $m_\sigma=0$. The first
term on the right hand side describes the variance of the inclusive
distribution and the Bose enhancement effect as we saw
in Section 3.  The additional terms which we discovered
in Section 4 could now be added to the right hand side.
It is of course the new, divergent term on which we shall focus
our attention.

\begin{figure}[t]
\centerline{\psfig{file=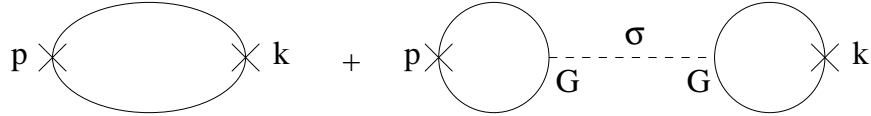,width=4.5in}}
\smallskip
\caption[]{Diagrammatic representation of the right hand side
of the correlator
(\ref{dndnsigma}). The crosses represent the insertions of $\Delta
n_p=:\pi_p\pi_p^*:$. The solid/dashed lines are the pion/sigma field
thermal propagators.}
\label{fig:diagrams}
\end{figure}

One can represent
both terms in this equation diagrammatically as in
Fig.~\ref{fig:diagrams}.  The singular term is due to the exchange of
the sigma in the process of forward pion-pion scattering. This
results in
a characteristic $1/m_\sigma^2$ singularity. A different way
of 
deriving the formula for the correlator would be to do a
straightforward diagrammatic expansion of $\langle{\Delta n_p\Delta
n_k}\rangle$. This will also allow one to 
include the effects of the nonzero
momentum modes of the sigma field. 
The
second term in (\ref{dndnsigma}) is the most singular one in this
correlator as $m_\sigma\to 0$ because it involves the exchange of the sigma 
field with zero momentum. We defer an analysis 
of less singular terms using the 
diagrammatic approach to future work.

\subsection{Application: Fluctuations of Mean $p_T$}

Using (\ref{dndnsigma}), we can
determine the fluctuations of any thermodynamic
observable which can be expressed in terms of the pion occupation
numbers. For a generic intensive observable $q=Q/N$ we find
\begin{eqnarray}
\langle{(\Delta q)^2} \rangle
&=& \left\langle{\left[\Delta\left(Q\over N\right)\right]^2 }\right\rangle
\nonumber\\
&=& {1\over \langle N\rangle^2}\left\{ 
\sum_p {v}^2_p (q_p-\langle q\rangle)^2 
+ {1\over V m_\sigma^2}{G^2\over T}
\left[\sum_p{{v}^2_p\over\omega_p}
(q_p-\langle q\rangle)\right]^2\right\}\ .
\label{qmsigma}
\end{eqnarray}
where we have displayed the factor
of $V$ explicitly to show that both terms are
of the same order in $V$.  
(Recall that $\sum_p\rightarrow V\int d^3p/(2\pi)^3$.) 
As before, the intensive observable of primary interest will be
the mean transverse momentum in an event, $q=p_T$. 
It is clear from
(\ref{qmsigma}) that the fluctuations of $p_T$ increase
near the critical point and 
diverge at the
critical point, where $m_\sigma$ vanishes.
We will give a  quantitative estimate of the effect in 
Section~\ref{sec:effsize}. We must first estimate the
size of the coupling $G$ and the value 
of $m_\sigma$ near the critical point.\footnote{Similarly, 
one can also calculate the fluctuations of $\tilde T$
defined in Section~\ref{sec:ttilde} and show that these fluctuations also
increase near the critical point and diverge at the critical point.}

\subsection{The Size of the Coupling $G$}

The strength of the singular contribution to the pion correlator
near the critical point depends on the size of the coupling $G$ 
between the $\sigma$ and $\pi$:
\begin{equation}
{\cal L}_I = G \, \sigma\, \pi_i\pi_i
\end{equation}
where the isospin index $i=1,2,3$ is summed.
We first make a phenomenological
estimate of the magnitude of $G$ in
vacuum, and then estimate by how much $G$ is reduced
near the critical point E. 

The value of this coupling in the vacuum can be inferred independently
from two considerations: (i) from the relationship between the sigma and
pion masses and $f_\pi$; (ii) more directly, from the width of the
sigma. We shall use both and compare.

One way of estimating the vacuum value of $G$ is 
to use the Gell-Mann -- L\'evy linear sigma model\cite{GellmannLevy},
in which the Lagrangian describing the dynamics of the
four component field $\phi_\alpha = (\phi_0,\mbox{\boldmath $\pi$})$
is given by
\begin{equation}
{\cal L} = \int d^4 x \left\{ \frac{1}{2} \partial^\mu \phi_\alpha
\partial_\mu \phi_\alpha - \frac{\lambda}{4}\left(\phi_\alpha\phi_\alpha
-v^2\right)^2 + H \phi_0 \right\}\ ,
\label{sigmamodel}
\end{equation}
where the O(4)-breaking field $H$ is proportional to the quark mass:
$H=m\langle\bar\psi\psi\rangle/f_\pi$. 
The vacuum expectation 
value $\langle{\phi_0}\rangle$ is nonzero and should be set
equal to $f_\pi$.
The $\sigma$ field is then defined by 
$\sigma = \phi_0 - \langle{\phi_0}\rangle$. 
Setting $f_\pi=93$ MeV, 
$m_\pi=135$ MeV and $m_\sigma = 600$ MeV 
fixes all three parameters in the potential,
and in particular yields $\lambda=20.0$.
If we rewrite (\ref{sigmamodel}) in terms of $\sigma$, we
find a term $\lambda f_\pi \sigma \mbox{\boldmath $\pi$}^2$, from
which we conclude that 
\begin{equation}
G=\lambda f_\pi \sim 1900\ {\rm MeV}\ .
\end{equation}

This value for $G$ seems large at first sight, but such
a large coupling is in fact required by experiment. 
In order to see this, 
we evaluate the width of the sigma due to its tree-level decay
into two pions, and find
\be
\Gamma = \frac{3 G^2}{8\pi}\frac{p}{m_\sigma^2}
 = \frac{3 G^2}{8\pi m_\sigma^2}
\sqrt{\left(\frac{m_\sigma}{2}\right)^2-m_\pi^2}\label{vacuumwidth}
\sim 300\ {\rm MeV}\ ,
\ee
where we have 
used $m_\sigma=600$ MeV as above.
The width of the sigma is known experimentally to be so large
that this ``particle'' is only seen as a broad bump in the $s$-wave
$\pi$-$\pi$ scattering cross-section. An estimate of $300$ MeV
for this width is therefore reasonable. We conclude that 
the vacuum $\sigma \pi\pi$ coupling must be at least as
large as $G\sim 1900$ MeV, since the sigma would otherwise
be too narrow.

Our estimate makes it clear that the vacuum value of $G$ 
would not change much if one were to take the chiral limit
$m\rightarrow 0$. The situation is different at the critical
point.  Taking the quark mass to zero while following
the critical endpoint leads one to the tricritical point P
in the phase diagram for QCD with two massless quarks.
At this point, $G$ vanishes as we discuss below.
This suggests that at E, the coupling $G$ is less than in vacuum.
Our goal in the remainder of this subsection is to 
use what we know about physics near the tricritical point P
to make an estimate of how much the coupling $G$ is 
reduced at the critical endpoint E (with the quark mass
$m$ having its physical value), relative to the vacuum value 
$G\sim 1900$ MeV
estimated above.

We begin by recalling some known results. (For details,
see Refs. \cite{BeRa97,HaJa97,SRS}.)
In QCD with two massless quarks, a spontaneously broken
chiral symmetry is restored at finite temperature. This
transition is likely second order and belongs in the universality
class of $O(4)$ magnets in three dimensions.  At zero $T$,
various models suggest that the chiral symmetry restoration
transition at finite $\mu$ is first order. Assuming that
this is the case, one can easily argue that there 
must be a tricritical point P in the $T\mu$ phase
diagram, where the transition changes from first
order (at higher $\mu$ than P) 
to second order (at lower $\mu$), and such a tricritical
point has been found in a variety of models.\cite{BeRa97,HaJa97,italians}
The nature of this point can be understood by considering
the Landau-Ginzburg effective potential for $\phi_\alpha$,
order parameter of chiral symmetry breaking:
\begin{equation}\label{phi6potential}
\Omega(\phi_\alpha) = \frac a2 \phi_\alpha\phi_\alpha 
+ \frac b4 (\phi_\alpha\phi_\alpha)^2 + \frac c6 (\phi_\alpha\phi_\alpha)^3\ .
\end{equation}
The coefficients $a$, $b$ and $c>0$ are functions of $\mu$
and $T$. The second order phase transition line described by
$a=0$ at $b>0$ becomes first order when $b$ changes sign, 
and the tricritical point P is therefore the point at which $a=b=0$.
The critical properties of this point can be inferred from
universality\cite{BeRa97,HaJa97}, and the exponents are as
in the mean field theory (\ref{phi6potential}). We will
use this below.  Most important in the present
context is the fact that
because $\langle \phi\rangle=0$ at P, there is 
no $\sigma\pi\pi$ coupling, and $G=0$ there.

In real QCD with nonzero quark masses, the second order phase transition
becomes a smooth crossover and the tricritical point P becomes E, the 
second order critical endpoint of a first order phase
transition line. Whereas at P there are four massless 
scalar fields undergoing critical long wavelength fluctuations,
the $\sigma$ is the only field which becomes massless
at the point E, and the point E is therefore in the Ising
universality class\cite{BeRa97,HaJa97}. The pions remain
massive at E because of the explicit chiral symmetry
breaking introduced by the quark mass $m$.  Thus, 
when we discuss physics near E as a function of $\mu$ and
$T$, but at fixed $m$, we will use universal scaling relations
with exponents from the three dimensional Ising model.  
Our present purpose, however, is to imagine varying $m$
while changing $T$ and $\mu$ in such a way as to stay
at the critical point E, and ask how large $G$ (and $m_\pi$)
become once $m$ is increased from zero (the tricritical
point P at which $G=m_\pi=0$) to its physical value.
For this task, we use exponents describing universal
physics near P.  Applying tricritical scaling relations
all the way up to a quark mass which is large
enough that $m_\pi$ is not small compared to $T_c$ may
introduce some uncertainty into our estimate.

In order to determine the trajectory of 
the critical line
of Ising critical points E as a function of quark 
mass $m$,\footnote{See Ref. \cite{rajreview} for a derivation
of the analogous line of Ising points emerging from the tricritical
point in the QCD phase diagram at zero $\mu$ as a function of
$m$ and the strange quark mass $m_s$. This tricritical point can
be related to the one we are discussing by varying $m_s$\cite{SRS}.}
it is sufficient to consider the effective potential 
only as a function of the single component $\phi_0\equiv\phi$
of the four-component order parameter.
When the quark mass is nonzero
we can add terms containing odd powers 
of $\phi$: $\phi^3$ and $\phi^5$, in addition to just $\phi$.
We shall assume that the linear term provides the leading effect,
and check this assumption for self-consistency a posteriori.
So, we have at nonzero $m$
\begin{equation}\label{omegam}
\Omega(\phi) = -m\phi + \frac a2 \phi^2 + \frac b4  \phi^4 
+ \frac c6 \phi^6.
\end{equation}
We assume that the units of mass are chosen in such a way that the 
coefficient of the 
linear term in (\ref{omegam}) assumes this simple form. That 
is, instead of writing it as $H=m M^2$, we are 
using units with $M=1$. Stable or metastable thermodynamic
phases are described by 
minima of $\Omega$, at which $\Omega'=0$.
At the critical point E, $\Omega'=0$ and in addition both 
$\Omega''$ and $\Omega'''$ 
vanish.
This is because three roots
of the polynomial $\Omega'(\phi)$ coalesce (two minima of $\Omega$
and one maximum in between). So, we have three conditions:
\begin{equation}\label{o'}
\Omega' = - m + a\langle\phi\rangle + b\langle\phi\rangle^3 + c\langle\phi\rangle^5 = 0;
\end{equation}
\begin{equation}\label{o''}
\Omega'' = a + 3b\langle\phi\rangle^2 + 5c\langle\phi\rangle^4 = 0;
\end{equation}
\begin{equation}\label{o'''}
\Omega''' = +6b\langle\phi\rangle + 20c\langle\phi\rangle^3 = 0.
\end{equation}
These conditions allow us to express  
$a$, $b$ and $\langle\phi\rangle$ (the value of $\phi$ at the minimum),
as functions of $m$ and $c$.  We neglect any change in $c$;
it is the vanishing of $m$ at P which is of interest to us.
Solving these equations by working up from the last to the first
and keeping only the exponents (neglecting pre-factors) we find:
\begin{equation}\label{mpow}
a \sim m^{4/5}; \qquad -b\sim m^{2/5}; \qquad \langle\phi\rangle\sim m^{1/5}.
\end{equation}
The power $1/5$ is easy to understand: it is $1/\delta$, where
$\delta=5$ for the $\phi^6$ potential.

At $m=0$, the tricritical point P at $a=b=0$ has $\langle\phi\rangle=0$;
the expressions (\ref{mpow}) describe how 
the location of the critical point E in the $ab$ plane, and
the value of $\langle\phi\rangle$ at E, change
as $m$ is increased from zero.  From these, we will determine
how $m_\pi$ and $G$ at E vary with $m$, after two asides.  First, note 
that from these universal arguments
we learn nothing about the location of the tricritical point $a=b=0$
in the $T\mu$ plane.  One can only make rather crude 
estimates of the position of this point, as we have done
in Ref. \cite{SRS}.
Our
main purpose here and in \cite{SRS} is to 
tell experimentalists
{\it how} to find P, so that they can find it 
and tell us {\it where} it is.
Second, we must
estimate the size of the $\phi^3$ and $\phi^5$ terms we
have neglected. Assuming that both terms come with coefficients which are at
least linear in $m$ (higher odd powers of $m$ are possible, but will
make the size of these terms even smaller) and using the $m$ power
counting of (\ref{mpow}), we see that while all the terms in
(\ref{omegam}) are ${\cal O}(m^{6/5})$, the $\phi^3$ and $\phi^5$
terms contribute at most ${\cal O}(m^{8/5})$ and ${\cal O}(m^2)$
respectively.

To follow $m_\pi$ and $G$, 
we need the full dependence of $\Omega$ 
on the $\phi_0$ and {\boldmath $\pi$}
fields:
\begin{equation}\label{osigpi}
\Omega(\phi,\mbox{\boldmath $\pi$} ) =  -m\phi 
+ \frac a2 (\phi^2 + \mbox{\boldmath $\pi$}^2)
+ \frac b4  (\phi^2 + \mbox{\boldmath $\pi$}^2)^2 
+ \frac c6 (\phi^2 + \mbox{\boldmath $\pi$}^2)^3\ .
\end{equation}
For the pion mass, we need to expand around $\phi=\langle\phi\rangle$ and
$\mbox{\boldmath $\pi$}=0$ and collect order $\mbox{\boldmath $\pi$}^2$ terms:
\begin{equation}
\Omega = {\mbox{\boldmath $\pi$}^2\over2}(a+b\langle\phi\rangle^2
+c\langle\phi\rangle^4)
+ \ldots\ .
\end{equation}
We can now read off the pion mass: 
\be 
m_\pi^2 =
a+b\langle\phi\rangle^2+c\langle\phi\rangle^4, 
\ee
which, according to (\ref{o'}), means: $m_\pi^2 = m/\langle\phi\rangle$.
Using (\ref{mpow}) we find:
\begin{equation}\label{mpim}
m_\pi^2 \sim m^{4/5}.
\end{equation}
Assuming that the dimensionful factor in this formula is of the same
order of magnitude as the one in the zero $T$ and $\mu$ formula
$m_\pi^2\sim m$ we conclude that the pion mass does not change
much from its vacuum value, and is likely to be very slightly bigger
(by a factor of order $(\Lambda_{\rm QCD}/m)^{1/10}$). 
This is similar to what is known to happen near $T_c$ for $\mu=0$.
(See Ref. \cite{rajreview} for a review.)

To determine the constant $G$, we need to collect the 
$\sigma\mbox{\boldmath $\pi$}^2$ 
terms in
(\ref{osigpi}) where, as before, $\sigma=\phi-\langle\phi\rangle$.  Only the
last two terms contribute and we find
\begin{equation}
\Omega = \sigma\mbox{\boldmath $\pi$}^2
(b\langle\phi\rangle + 2c\langle\phi\rangle^3) + \ldots\ .
\end{equation}
This means $G= b\langle\phi\rangle + 2c\langle\phi\rangle^3$ 
which, according to
(\ref{o'''}), gives $G=2b\langle\phi\rangle/5=-4c\langle\phi\rangle^3/3$. 
Using the $m$ power counting
(\ref{mpow}) we find
\begin{equation}\label{Gm}
G\sim m^{3/5}.
\end{equation}
Thus the coupling $G$ is suppressed
compared to its ``natural'' vacuum value $G_{\rm vac}$ by
a factor of order $(m/\Lambda_{\rm QCD})^{3/5}$.
Taking $\Lambda_{\rm QCD}\sim 200$ MeV, $m\sim 10$ MeV we
obtain our estimate 
\begin{equation}
G_E\sim \frac{G_{\rm vac}}{6} \sim 300 {\rm ~MeV}\ .
\end{equation}
The main 
source of uncertainty in this estimate 
is our inability
to compute the various nonuniversal
masses which enter the estimate as prefactors in front
of the $m$ dependence which we have followed. In other words,
we do not know the correct value to use for 
$\Lambda_{\rm QCD}$ in the suppression factor 
which we write as $(m/\Lambda_{\rm QCD})^{3/5}$.


\subsection{Finite Size and Finite Time Effects}
\label{sec:finsiz}

The final ingredient needed for the estimate of 
the size of the effect described in Sections 5.1 and 5.2
is an estimate of $m_\sigma$.  We found that 
$\langle \Delta n_p \Delta n_k\rangle$ is infinite
when $m_\sigma=0$, at E.  This singularity occurs
because the correlation length $\xi$ of the sigma field
is infinite.  In practice, however, there are important
restrictions on how large $\xi$ can become. The fireball
created in a heavy ion collision has a finite size and 
lives for a finite time; both restrict $\xi$.  Similar
considerations affect the estimate of the size 
of the effect described in Section~\ref{sec:corrCACB}.
There, we found an anti-correlation in $\langle \Delta n_p \Delta n_k\rangle$
which vanishes as the specific heat of the system diverges.
The limit on $\xi$ introduced by finite size and finite
time effects also limits how large the heat 
capacity $C_V$ becomes.

We discuss finite size scaling first. If the system is
infinite, a singular thermodynamic quantity such as $C_V$ 
diverges at the critical point.  If the system is large
relative to microscopic scales ($\sim 1$ fm in our case)
but finite, then $C_V$ exhibits a peak at the critical
point which becomes narrower and higher as larger
and larger systems are considered. Finite size
scaling analysis tells us how the magnitude 
of the peak scales with the system size.
The scaling postulate tells us that
the singular parts of all observables are due to the diverging
correlation length $\xi$ and can be characterized by an appropriate
critical index: $Q_{\rm sing} \sim \xi^{\Delta_Q}$, where
$Q$ could be $C_V$ or some other quantity which diverges
at E. In a finite
system the growth of the correlation length is limited by the
size of the system: $\xi_{\rm max} \sim R$. Therefore, the
magnitude of the singularity in a given thermodynamic quantity 
(the height of the peak) depends on the size of the system as
\begin{equation}\label{finsize}
Q_{\rm max} \sim R^{\Delta_Q}.
\end{equation}

Similarly, if the system is not allowed enough time to equilibrate,
the singularity is also smeared. The magnitude of the singularity in
this case can
be estimated using finite time, or dynamic, 
scaling~\cite{HoHa}. In this case
the scaling postulate tells us that the typical equilibration time
diverges at the critical point (critical slowing down), with
this divergence related to that of the correlation length
by $t\sim\xi^{\Delta_t}$.   
Reversing this relation,
we conclude that if the typical time allowed for the system to
equilibrate is limited to $\tau$, the correlation length 
can only grow up to $\xi_{\rm max} \sim \tau^{1/\Delta_t}$.
Thus, in this case
\begin{equation}\label{fintime}
Q_{\rm max} \sim \tau^{\Delta_Q/\Delta_t}.
\end{equation}

The calculation of the numerical prefactors in (\ref{finsize})
and (\ref{fintime}) requires
precise knowledge of the QCD dynamics and is not feasible at this
time. The exponents, however, are universal and can be understood
by relating them to suitable exponents in the three-dimensional
Ising model.
For example, the exponent for the specific heat
$C_V$ at the end point E was determined in~\cite{SRS}:
\begin{equation}\label{DCV}
\Delta_{C_V} =
\left(\gamma\over\nu\right)_{\rm 3d-Ising} \approx 2.
\end{equation}
$\Delta_{C_V}$ is {\it not} given by the (smaller)
exponent $(\alpha/\nu)_{\rm 3d-Ising}$ because of the obliqueness of the first
order phase transition line relative to the $T$ axis on the phase
diagram as explained in~\cite{SRS}. The idea is 
that at the critical point, 
$C_V=\partial^2\Omega/\partial T^2$ is related to some linear
combination of the Ising model susceptibilities 
$\partial^2\Omega/\partial t^2$, 
$\partial^2\Omega/\partial t\partial h$,  and $\partial^2\Omega/\partial h^2$
where the Ising model temperature axis $t$ and magnetic field
axis $h$ are oblique relative to the $T$ and $\mu$ axes. $C_V$
is controlled by the most divergent of the three Ising model
susceptibilities, which is $\partial^2\Omega/\partial h^2$,
and (\ref{DCV}) results.

The dynamic scaling exponent $\Delta_t$, which 
is often called $z$,  is also
universal. 
The dynamic universality class
of a system is sensitive to 
details of the dynamics such as whether the order parameter
is or is not conserved and whether the system has other
conserved quantities.
The determination of $\Delta_t$ is a rather involved problem in some
cases~\cite{HoHa}. If we assume
that QCD at the critical point E falls
into the dynamic universality class of the gas-liquid phase transition
(model H in the classification of Hohenberg 
and Halperin\cite{HoHa}) the exponent $\Delta_t$ can be
estimated as $\Delta_t\approx 3$. It may therefore turn 
out that because $\Delta_t>1$ the finite time scaling
restriction (\ref{fintime}) may be somewhat more restrictive
in a heavy ion collision than the finite size scaling restriction
(\ref{finsize}).

Let us estimate some typical numbers
for central PbPb collisions at the SPS.
We start with  an estimate for the relevant size in the longitudinal
and transverse directions beyond which $\xi$ cannot grow.
The  longitudinal expansion extends the 
longitudinal size of the fireball considerably, but
regions with different rapidities freeze out at different times,
and a homogeneous freeze-out at a single freeze-out time for all rapidities
is not a good approximation. A similar (although not identical)
problem has already been faced in two pion interferometry, which
provides sizes (and durations) not of the whole system, but
of a ``patch'' large enough that particles emitted from it
can still interfere.  
The size of such patches depends
somewhat on the direction and magnitude of the total momentum
of the pion pair used in the interferometric measurement, but is
an approximate measure of the size of the system over which
freeze-out is homogeneous.
The longitudinal size of such a patch for central PbPb collisions 
at 160 AGeV is
estimated to be\cite{HeinzJacak}  
$2R_L\approx 12-14$ fm. At its ends, the rapidity
difference is already about 1. 
We therefore estimate that sigma field correlation length 
is limited by finite size effects to be less than 
$2R_L$.

The size $D_T$ in the transverse direction beyond which $\xi$ cannot
grow
can be estimated in two ways. The initial   size
is that of the diameter of the nuclei,  $D_{\rm Pb}\sim 14$ fm. 
The transverse  (radial) flow makes the physical size of the
freeze-out surface larger than the nuclear radius, by 30-50\% at freeze-out. 
Therefore, it must be the case that $D_T<20$ fm.  This is,
however, an overestimate.
Because of the relativistic transverse expansion, regions with different
positions in the transverse directions cross the
transition region and then freeze-out at different
times. Therefore, as for $R_L$ above we can use the size
of ``patches'' observed via two particle interferometry as a guide, 
the sigma field correlation length in the transverse
directions to be less than
$D_T=2R_T\approx 10-12$ fm\cite{HeinzJacak}.
It therefore seems that the relevant longitudinal
and transverse length scales at freeze-out are about the same,
and we conclude that based on finite size restrictions alone 
\be \label{xifinsize}
{\xi\over l} < 12 
\ee 
where $l$ is the ``microscopic length'' of order 1 fm.

We now turn to the restriction on the correlation
length which arises from the fact that the matter
created in a  heavy ion
system does not enjoy an infinite period of time
in which to equilibrate.
The expansion time can be defined through the corresponding
``Hubble constant''
\be\label{hubbleconstant}
H={1 \over \tau_H} = {ds \over s dt}.
\ee
where $s$ is entropy density. We use the entropy density
in the definition because the total entropy is conserved 
during adiabatic expansion, and we are assuming that by the time
the system is traversing the transition region and then
freezing out, the expansion can be treated as adiabatic.
Hydrodynamic models\cite{HS} suggest that at SPS energies,
heavy ion collisions have 
$\tau_H \approx 10-20$ fm$/c$.
If we simply use this value of $\tau_H$ and
neglect dimensionless factors in the scaling relation
(\ref{fintime}) we would find
\be\label{xifintime}
{\xi\over l} < \left(\tau_H\over l\right)^{1/\Delta_t} \sim 2.5. 
\ee
In spite of the long expansion time, the relatively large value of
the dynamical exponent $\Delta_t$ can make the finite time restriction more 
severe than the finite size one (\ref{xifinsize}).
In other words,
although the size of the system allows the correlation length to
become as large as 12 fm, there may 
not be enough time for such long wavelength 
fluctuations
to develop due to critical slowing down. 
The estimate (\ref{xifintime}) is suspect for several reasons.
First, there may be a large dimensionless proportionality
constant in this relation which is unknown to us. (In contrast,
the finite size estimate (\ref{xifinsize}) is a consequence
of geometry, and unless the homogeneous region is larger
than we estimate, it is unlikely that the finite size
bound on $\xi$ can be evaded.)  Second, in making
the estimate (\ref{xifintime}) we have estimated the
``bad'' effect of critical slowing down, namely the
fact that a 12 fm correlation length will take longer
than 12 fm to develop, but we have not taken into
account a compensating ``good'' effect of critical
slowing down:  because of the large specific heat,
the system will spend an unusually long time with
a temperature in the vicinity of the critical point.
Because of the uncertainties in (\ref{xifintime}),
we shall use  $\xi_{\rm max} \sim 6$ fm as a rough estimate
of the largest correlation length possible if control
parameters are chosen in such a way that the system
freezes out close to the critical point. More
detailed study of the time evolution of the temperature
of the system, of the appropriate choice for $\tau_H$, 
and of the dimensionless factors in (\ref{xifintime})
are required in order to properly estimate whether
finite time effects restrict $\xi_{\rm max}$ further.


Since the thermal contact effect discussed in Section 4 depends
on the divergent heat capacity $C_A$, we need to estimate
how large $C_A$ can get, given the finiteness of the system.
Using the exponent (\ref{DCV}), we can estimate
the ratio of the maximum value of that part of $C_A$ 
which would be singular in an infinite system to the
``normal'' value of $C_A$ for a degree of freedom
with a correlation length $l\sim 1$ fm as
\begin{equation}
{(C_A)^\sigma_{\rm max} \over (C_A)_{\rm norm}} 
\sim \left(\xi_{\rm max}\over l\right)^{\Delta_{C_V}}
\sim 36\ .
\end{equation}
This does not mean that the specific heat $C_A$ is multiplied
by 36, because it receives a nonsingular contribution from
other degrees of freedom. However, it suggests that 
in using Eq. (\ref{FTresult}) to estimate how much
the anti-correlation induced reduction of $F_T$ is
weakened at the critical point, it is reasonable
to expect that 
$C_A$ can be up to an order of magnitude larger there than
it is near $T_c$ far from the critical point.


We now return to our discussion of the effects of the 
long wavelength sigma fluctuations on the fluctuations
of the pions, encoded in the microscopic correlator (\ref{dndnsigma}).
We derived (\ref{dndnsigma}) using mean field theory, 
and would now like to discuss the effect of non-mean-field
corrections. We mentioned previously that fluctuations of
the sigma field around the minimum of $\Omega(\sigma)$ are not
small; we now argue that this does not make much difference
to the quantity of interest. One way to 
see how these corrections
can appear is to realize that, at higher order in $G$, diagrams
with $\sigma$-bubbles which are actually logarithmically
divergent as $m_\sigma\to 0$ will contribute. 
These bubbles have to be resummed and may
modify the exponent of the $m_\sigma^{-2}$ singularity
in~(\ref{dndnsigma}). 
This exponent is easy to infer from
universality arguments. 
Diagrammatically, the $1/m_\sigma^2$
is the zero momentum value of the sigma propagator, i.e., the
sigma field susceptibility.
For the 3d-Ising universality class we know the corresponding exponent
to be $\gamma/\nu=2-\eta$ which is $\approx2$ to 
within a few percent because $\eta$ is
small. We can therefore safely use the mean-field
formula~(\ref{dndnsigma}) with its $m_\sigma^{-2}$
divergence for our estimate, and will take 
$m_\sigma\sim 1/\xi_{\rm max}\sim 1/(6 {\rm ~fm})$.
It therefore turns out that even though the effects of
Section 4 depend on $C_V$ and the effects of
this Section depend on the sigma susceptibility,
both are controlled by the exponent $\gamma/\nu\approx 2$.

\subsection{Magnitude of the Effects}
\label{sec:effsize}

We now have all the ingredients in place for our estimate of the
size of the effect of the critical fluctuations 
of the sigma field on the
fluctuations of the observed pions, via the coupling $G$.
We reproduce here Eq. (\ref{qmsigma})
\begin{eqnarray}
\langle{(\Delta q)^2} \rangle
&=& \left\langle{\left[\Delta\left(Q\over N\right)\right]^2 }\right\rangle
\nonumber\\
&=& {1\over \langle N\rangle^2}\left\{ 
\sum_p {v}^2_p (q_p-\langle q\rangle)^2 
+ {1\over V m_\sigma^2}{G^2\over T}
\left[\sum_p{{v}^2_p\over\omega_p}
(q_p-\langle q\rangle)\right]^2\right\}\ ,
\label{qmsigma2}
\end{eqnarray}
which we now apply for $q=p_T$. We have restored the factor of $V$.
The first term in the curly brackets includes the single particle
inclusive distribution enhanced by the Bose effect. The
second term is the effect we are interested in now.
As we did in our estimate of the effects of 
energy conservation and thermal contact in
Section~\ref{sec:thermalcontactflucts}, 
we shall express the size of the effect of interest 
as the
ratio of the entire expression in curly brackets in (\ref{qmsigma2})
to the first term in these brackets. We find
\begin{equation}\label{FTresultmu60}
F_\sigma = 1 + 0.35 \left(G_{\rm freeze-out}\over 300\ {\rm MeV}\right)^2 
\left(\xi_{\rm freeze-out}\over 6\ {\rm fm}\right)^2 
\qquad{\rm for~} \mu_\pi = 60 {\rm MeV}
\end{equation}
and 
\begin{equation}\label{FTresultmu0}
F_\sigma = 1 + 0.14 \left(G_{\rm freeze-out}\over 300\ {\rm MeV}\right)^2 
\left(\xi_{\rm freeze-out}\over 6\ {\rm fm}\right)^2 
\qquad{\rm for~} \mu_\pi = 0\ ,
\end{equation}
where we have taken $T=120$ MeV. As in Section 4, the effect will
be diluted by about a factor of two 
because not all of the pions which
are observed are direct.
We have written the estimates (\ref{FTresultmu60},\ref{FTresultmu0})
in such a way that the largest uncertainties are manifest.
The size of the effect depends quadratically on the coupling $G$.
We argued above that $G$ is reduced
to $G_E\sim 300$ MeV at the critical point but, as we explained,
there are caveats in this argument. Furthermore, freeze-out
may occur somewhat away from the critical point, in which
case $G$ would be somewhat larger, although still much
smaller than its vacuum value.  
The size of the effect also depends quadratically
on the sigma correlation length at freeze-out, and we have
seen that there are many caveats in an estimate like 
$\xi_{\rm freeze-out}\sim\xi_{\rm max}\sim 6$ fm. 
Finally, the effect is sensitive to the value of $\mu_\pi$.
There are reasons to believe that $\mu_\pi$ may be smaller
near the critical point than far from it.  Recall that 
$\mu_\pi$ is zero at chemical freeze-out, and then grows
until thermal freeze-out.  At the critical point, the transition
temperature $T_c$ is somewhat lower than at lower baryon 
chemical potential $\mu$, and this suggests that $T_{\rm ch}$
may be lower than the value measured in 160 AGeV collisions.  
Furthermore, we have argued in Ref. \cite{SRS} that the thermal
freeze-out temperature $T_{\rm f}$ will be somewhat higher
in the vicinity of the critical point, because the system
lingers there and expands for a while with a temperature 
near $T_c$.\footnote{The expansion rate $H$ of (\ref{hubbleconstant})
does not decrease. However, the rate of change of $T$ with
time is reduced because of the large specific heat.}
If the temperature window $T_{\rm ch}-T_{\rm f}$ is small near
the critical point, then $\mu_\pi$ may be significantly smaller
than $60$ MeV there.

We have studied two different effects on 
$\sqrt{F}$
in Sections 4 and 5.  The effects of energy conservation and thermal contact
lead to a few percent reduction in 
$\sqrt{F}$;
this reduction will be significantly lessened near the critical point
due to the divergence of $C_A$.  This yields perhaps a $1-2\%$
increase in $\sqrt{F}$
near the critical point.
The direct effect of
the critical fluctuations which we have estimated in
this section is a further, larger, 
increase in 
$\sqrt{F}$
by a factor of $\sqrt{F_\sigma}$. We have displayed the
various uncertainties in the factors contributing to our estimates 
(\ref{FTresultmu60},\ref{FTresultmu0}) so that when
an experimental detection of an increase 
and then subsequent decrease in 
$\sqrt{F}$
occurs, as control parameters are varied and the critical point
is approached and then passed, we will be able to use 
the measured magnitude of this nonmonotonic effect to constrain
these uncertainties.  It should already be clear that an 
effect as large as $10\%$ in $\sqrt{F_\sigma}$ is easily possible; 
this would be 50 times larger than the statistical error in
the present data.

Once we have evaluated the microscopic correlator
$\langle \Delta n_p \Delta n_k \rangle$, we can estimate
the effect of the critical fluctuations of the sigma 
field on
the fluctuations of many different pion observables.
To this point, we have focussed on $\langle (\Delta p_T)^2\rangle$;
we now give a brief account of the effect
on $\langle (\Delta N)^2\rangle$ and $\langle \Delta N \Delta p_T\rangle$.
We can use the microscopic correlator (\ref{dndnsigma}) to obtain
\begin{eqnarray}
\langle (\Delta N)^2\rangle &=&
\sum_p {v}^2_p  
+ {1\over V m_\sigma^2}{G^2\over T}
\left[\sum_p{{v}^2_p\over\omega_p}\right]^2 \nonumber\\
&\approx& \sum_p {v}^2_p\left\{1+
1.0\left(G_{\rm freeze-out}\over 300\ {\rm MeV}\right)^2 
\left(\xi_{\rm freeze-out}\over 6\ {\rm fm}\right)^2\right\} \ ,
\label{bigresult}
\end{eqnarray}
for $q=p_T$, $T=120$ MeV and $\mu_\pi=0$.
The coefficient which is 1.0 in the last line of (\ref{bigresult})
increases to 2.0 if $\mu_\pi=60$ MeV.  We see that
there can easily be a very large increase in the
multiplicity fluctuations of the direct pions 
near the critical point, due to the coupling 
between the direct pions and the critical fluctuations
of the sigma field.  As we have noted previously, 
the noncritical fluctuations of the total pion multiplicity
are dominated by the pions from resonance decay.
Using the result (\ref{multflucprediction}), we estimate
that the sigma induced critical multiplicity fluctuations 
of the direct pions can easily lead to a $10\%$ increase in 
the total $\langle (\Delta N)^2\rangle$.  This 
is comparable in magnitude to the effect on 
$\langle (\Delta p_T)^2\rangle$, and should be easily
detectable.  We will see in Section 6 that there should
be even a further increase in the fluctuations of 
the multiplicity of those pions with low $p_T$.

Turning now to the cross correlation between an intensive quantity
and $N$, we use the microscopic correlator
(\ref{dndnsigma}) to calculate
\begin{eqnarray}
\frac{\langle \Delta N \Delta q \rangle}{\langle q \rangle} &=&
{1\over \langle N\rangle\langle q\rangle} \left\{
\sum_p {n}^2_p (q_p-\langle q\rangle )
+ \frac{1}{V m_\sigma^2}\frac{G^2}{T}
\left[\sum_k\frac{{v}^2_k}{\omega_k}\right] 
\left[\sum_p\frac{{v}^2_p}{\omega_p}(q_p-\langle q\rangle)\right]
\right\}\nonumber\\
&\approx&
-0.021\left\{1+12.\left(G_{\rm freeze-out}\over 300\ {\rm MeV}\right)^2 
\left(\xi_{\rm freeze-out}\over 6\ {\rm fm}\right)^2\right\} 
{\rm ~for~} \mu_\pi = 0\nonumber\\
&\approx&
-0.046\left\{1+13.\left(G_{\rm freeze-out}\over 300\ {\rm MeV}\right)^2 
\left(\xi_{\rm freeze-out}\over 6\ {\rm fm}\right)^2 \right\}
{\rm ~for~} \mu_\pi = 60{\rm ~MeV},\label{dNdptsigma}
\end{eqnarray}
where we have taken $q=p_T$ and $T=120$ MeV.  This correlation
only receives contributions from nontrivial
effects, and we see that near the critical point, the contribution
from the interaction with the sigma field is dominant.
Once again, we see that this correlation is a very interesting
quantity to use to look for the critical point. As the critical
point is approached, thermal contact with a heat bath
whose heat capacity is diverging reduces the effect
of energy conservation, as seen in (\ref{dNdpTthermal}); 
we now see that this reduction is overcompensated by the 
larger increase
in the cross correlation which is
induced by the direct coupling of the pions
to the sigma field.  The lesson is clear: although
this correlation is small, it may 
increase in magnitude by a very large
factor near the critical point.


The effects of the critical fluctuations can be detected
in a number of ways. First, one can find
a nonmonotonic increase in  $F_\sigma$, the
suitably normalized increase in the variance of event-by-event
fluctuations of the mean transverse momentum.  
Second, one can find a nonmonotonic increase in 
$\langle (\Delta N)^2\rangle$. Both these effects
can easily be between one and two orders of magnitude 
greater than the statistical errors in present
data. Third, one can
find a nonmonotonic increase in the magnitude
of $\langle \Delta p_T\Delta N\rangle$.  This quantity
is small, and it has not yet been demonstrated that it
can be measured. However, it may change at the critical
point by a large factor, and is therefore worth measuring.
In addition to effects on these and many other observables,
it is perhaps most distinctive to measure the microscopic
correlator $\langle \Delta n_p \Delta n_k \rangle$ itself.
The term proportional to $1/m_\sigma^2$ in 
(\ref{dndnsigma}) has
a specific dependence on $p$ and $k$. It introduces off-diagonal correlations
in $pk$ space.  Like the off-diagonal anti-correlation discussed
in Section 4, this makes it easy to distinguish from the
Bose enhancement effect, which is diagonal in $pk$. 
Near the critical point, the off-diagonal anti-correlation
vanishes and the off-diagonal correlation due to sigma exchange grows.
Furthermore, the effect of $\sigma$ exchange is not restricted
to identical pions, and should be visible as correlations between
the fluctuations of $\pi^+$ and $\pi^-$.  The 
dominant diagonal term proportional to $\delta_{pk}$ in (\ref{dndnsigma})
will be absent in the correlator 
$\langle \Delta n^+_p \Delta n^-_k \rangle$, and the 
effects of
$\sigma$ exchange will be the dominant contribution to this quantity
near the critical point.

%% file: chap6.tex
\section{Pions From Sigma Decay}

There is another signature of freeze-out near 
the critical point discussed in~\cite{SRS} 
in addition to those we have analyzed
in depth above.  For choices of control parameters
such that freeze-out occurs at or near the critical endpoint, 
the excitations of the sigma field, sigma (quasi)particles, 
are nearly massless at freeze-out and
are therefore numerous.  Because the pions are massive
at the critical point, these $\sigma$'s cannot immediately
decay into two pions. Instead, they persist as the
temperature and density of the system further decrease.
During the expansion, the in-medium $\sigma$ mass rises
towards its vacuum value and eventually exceeds the 
two pion threshold.  Thereafter, the $\sigma$'s decay,
yielding a population of pions which do not get a chance
to thermalize because they are produced after freeze-out.
Here, we estimate the momentum spectrum of these pions
produced by delayed $\sigma$ decay. 
An event-by-event
analysis is not required in order to see these pions.
The excess multiplicity at low $p_T$ will appear and
then disappear in the single particle inclusive distribution
as control parameters are varied such
that the critical point is approached and then
passed.  

The event-by-event fluctuations of the multiplicity of these
pions reflect the
fluctuations of the sigma field whence
they came\cite{SRS}.  We estimate the resulting
increase in the event-by-event fluctuations of $N$, the total
pion multiplicity.

We begin with the inclusive single particle $p_T$-spectrum
of the pions from sigma decay.
We use the expression (\ref{vacuumwidth}) for the width
of the $\sigma$, but now treat $m_\sigma$ as time-dependent.
We should also take $G$ to evolve with time. However,
the dominant time-dependent effect is the opening
up of the phase space for the decay as
$m_\sigma$ increases with time and 
crosses the two-pion threshold. We will therefore treat
$G$ as a constant.  In Section 5, we estimated that
in vacuum with  $m_\sigma=600$ MeV,  the coupling is
$G\sim 1900$ MeV, whereas at the critical end point
with $m_\sigma = 0$, the coupling is reduced, perhaps
by as much as a factor of six or so.  In this section,
we need to estimate $G$ at the time when $m_\sigma$ is
at or just above twice the pion mass.  We will use
$G\sim 1000$ MeV, recognizing that we may be off by
as much as a factor of two.  

Let us parameterize
the time dependence of the sigma mass by
\begin{equation}
m_\sigma(t) = 2 m_\pi (1+ t/\tau)
\end{equation}
where we have defined $t=0$ to be the time at which
$m_\sigma$ has risen to $2m_\pi$ and 
have introduced the timescale $\tau$ over which
$m_\sigma$ increases from $2m_\pi$  to $4m_\pi$.
We will be interested in times $0<t < \tau$, for which
this linear parameterization of the
time dependence is not unreasonable.
Note that with this choice of notation, freeze-out
occurs at a negative time, and the collision begins at
an even more negative time.
Substituting into (\ref{vacuumwidth}), and working
to lowest order in $t/\tau$, we find
\begin{equation}
\Gamma(t)\sim \frac{3 G^2}{32\pi m_\pi}\sqrt{2 t/\tau} = D\sqrt{t/\tau}\ ,
\end{equation}
where we have defined
\begin{equation}
D=\frac{3\sqrt{2} G^2}{32\pi m_\pi} \sim (300\ {\rm MeV})\left(\frac{G}
{1000\ {\rm MeV}}\right)^2\ .
\label{cdef}
\end{equation}
$N(t)$, the number of $\sigma$'s present at time $t$, is
determined by 
\begin{equation}
\frac{1}{N(t)}\frac{dN(t)}{dt}= - \Gamma(t) = -D\sqrt{t/\tau}\ ,
\end{equation}
and is therefore
\begin{equation}
N(t)=N(0)\exp\left(-\frac{2}{3}\,D\,t^{3/2}\tau^{-1/2}\right)\ .
\label{Noft}
\end{equation}

We can now estimate the momentum distribution of the pions
produced in the decay of the sigmas, upon making the assumption
that the sigmas are at rest when they decay.  This is 
a good approximation for two reasons. First, as the system
expands after freeze-out, the sigma mass is increasing
as we have discussed. This means that the kinetic energy
of each sigma is decreasing.  Second, during
the time between freeze-out
and decay, some of the sigmas which happen 
to be moving outwards toward the less dense region of the
collision in which their mass would increase
more than allowed by energy conservation will 
instead be reflected
back inward.   
Each sigma which suffers such a reflection
loses momentum, as the reflection occurs as if off an outward moving
surface.  This effect confines the sigmas to the densest region
of the plasma, where their mass remains low for the longest time,
and in addition reduces their momenta.  We do not attempt a quantitative
estimate of these two momentum-reducing effects here. Suffice
to say that since at freeze-out the typical sigmas will have momenta of
order the freeze-out temperature or less, we think it reasonable
to approximate them as being at rest at time $t=0$ when
they begin to decay.

Sigmas which decay at rest at time $t$ each yield two pions with
momenta $p\sim m_\pi\sqrt{2t/\tau}$, to lowest order in
$t/\tau$.  As a result, the number of pions with momenta
$m_\pi\sqrt{2t/\tau}<p<m_\pi\sqrt{2(t+dt)/\tau}$ is 
$-2dt (dN(t)/dt)$ with $N(t)$ given by (\ref{Noft}).
Upon making suitable substitutions, we find that
the number of pions with momenta between $p$ and $dp$
is
\begin{equation}
dN = \sqrt{2}\, N(0) \, D \,\tau \,\frac{p^2 dp}{m_\pi^3} 
\exp\left(-\frac{1}{3\sqrt{2}}\,D\,\tau\,
\frac{p^3}{m_\pi^3}\right)\ .
\label{momentumdist}
\end{equation}
With the momentum distribution in hand, we determine
the mean pion momentum to be
\begin{equation}
\overline{p}^{\rm inc} = 2^{1/6} 3^{-2/3} \Gamma(1/3) \,m_\pi (D\tau)^{-1/3}
=1.45 \,m_\pi (D\tau)^{-1/3}\ .
\end{equation}
Large $\tau$ corresponds to slow expansion and a sigma mass
which consequently increases only slowly with time; 
large $D$ corresponds to a large coupling constant $G$.
It therefore makes sense that if $D\tau$ is large, the sigmas decay 
before the sigma mass has increased far above threshold,
and the resulting pions have small momenta.
We defined $\tau$ to be the time it takes the $m_\sigma$
to increase from $2m_\pi$ to $4m_\pi$.  This timescale is hard to
estimate, but our result is not strongly dependent on $\tau$.
It seems likely that $5\ {\rm fm}<\tau<20 \ {\rm fm}$
and we therefore quote our result as
\begin{equation}
\overline{p}^{\rm inc} \sim 0.58 
\,m_\pi \left(\frac{1000\ {\rm MeV}}{G}\right)^{2/3}
\left(\frac{10\ {\rm fm}}{\tau}\right)^{1/3}\ ,
\end{equation}
where we have used (\ref{cdef}).
We therefore estimate that if freeze-out occurs near the critical
point, there will be a nonthermal population of pions
with transverse momenta of order half the pion mass  
distributed according to (\ref{momentumdist}).

How many such pions can we expect?  That is, how large is $N(0)$?
This is determined by the sigma mass at freeze-out.  If 
$m_\sigma$ is comparable to $m_\pi$ at freeze-out, then there are half as many
$\sigma$'s at freeze-out as there are charged pions.  Since each
sigma decays into two pions, and two thirds of those pions
are charged, the result is that the number of charged pions
produced by sigma decays after freeze-out is $2/3$ of
the number of charged pions produced directly by the freeze-out
of the thermal pion gas.   Of course, if freeze-out occurs closer
to the critical point 
at which $m_\sigma$ can be as small as $(6 {\rm ~fm})^{-1}$, there would
be even more sigmas.  We therefore suggest that as experimenters
vary the collision energy, one way they can discover the critical
point is to see the appearance and then disappearance of a population
of 
pions with $\langle p_T\rangle \sim m_\pi/2$ which are almost as numerous
as the direct pions.
Yet again, it is the nonmonotonicity of this signature as
a function of control parameters which makes it distinctive.

As we discussed briefly in \cite{SRS}, the event-by-event
fluctuations in the multiplicity of these low momentum pions
are also of interest.  If we were able to measure
the multiplicity of sigma quasiparticles at freezeout, 
we would find fluctuations given by
\begin{equation}\label{multflucsigma}
\langle (\Delta N^\sigma)^2\rangle = \sum_p \langle n_p^\sigma\rangle
\left( 1 + \langle n_p^\sigma\rangle \right) \ ,
\end{equation}
where the $n_p^\sigma$'s are the sigma occupation numbers.  
In the present analysis, we neglect the effects of
interactions among the sigmas and just take $n_p^\sigma$ 
as for an ideal Bose gas with small $m_\sigma$. We expect
that this makes our prediction for the fluctuations an underestimate.
Since $m_\sigma$ is small, 
the low momentum modes have large occupation number, and have
fluctuations proportional to the square of their occupation number.
Each sigma eventually decays into two pions, whose momenta are
determined by the time at which the sigma decays, rather
than by the sigma momentum at freeze-out.
It is therefore not
possible to make a measurement on the pions which 
restricts the $\sum_p$ in (\ref{multflucsigma}) to low $p_T$.
We therefore do the entire sum, and find that the variance of the
event-by-event distribution of the multiplicity of the 
$\sigma$-produced pions is
\begin{equation}
\langle (\Delta N)^2\rangle = 2 \langle N \rangle \left(
1+\overline{\langle n_p^\sigma\rangle}^{\rm inc}\right)
\end{equation}
where $N$ is the number of charged pions.
The factor of two arises because
every sigma which produces charged pions 
produces two charged pions, and was discussed in Section 3.
Taking $m_\sigma=0$ yields 
$\overline{\langle n_p^\sigma\rangle}^{\rm inc}\approx 0.37$,
and therefore 
\begin{equation}\label{masslessmultiplicityfluct}
\langle (\Delta N)^2\rangle \approx  2.74 \langle N \rangle\ .
\end{equation}
We have already seen in Section 5 that the critical fluctuations 
of the sigma field increase the fluctuations in the multiplicity
of the direct pions sufficiently that the 
increase in the fluctuation of the multiplicity of 
all the pions will be increased by about
$10\%$.  
We now see that in the vicinity of
the critical point, there will be a further nonmonotonic
rise in the fluctuations of the multiplicity of the 
population of pions with $\langle p_T\rangle
\sim m_\pi/2$ which are produced in
sigma decay.


%% file: sum.tex
\section{Summary and Outlook}

In order to estimate the magnitude of the effects of critical
fluctuations, one must first analyze the background, noncritical
fluctuations.  
NA49 data from PbPb collisions at 160 AGeV shows that the event-by-event 
distribution of the mean transverse momentum is as perfect a Gaussian
as the central limit theorem allows.  Since a system
in thermodynamic equilibrium exhibits Gaussian fluctuations, 
in Section 3 we give a quantitative answer to the question
of how much of the observed fluctuations are thermodynamic
in origin.  To this end, we model the matter at freeze-out
as an ideal gas of pions and resonances in thermal
equilibrium, estimate the resulting fluctuations, and compare
with the data.

We calculate the 
event-by-event fluctuations
of $p_T$, an intensive quantity
which is therefore little affected by nonthermodynamic fluctuations
in the initial size of the system.
We find that the resonances
turn out to be of little importance --- the
resonance gas prediction for 
$\langle (\Delta p_T)^2\rangle/\langle p_T\rangle^2$
is almost indistinguishable from that of an 
ideal Bose gas of pions. Furthermore, we have
verified quantitatively that the correlations between
pions introduced by the fact that some originate in
resonance decays can be neglected. We have computed
the effects of Bose enhancement, and find that they
increase $\langle (\Delta p_T)^2\rangle$ by only a few percent,
although the precision of the data should make effects
of this magnitude detectable.

The difficulty comes in the treatment of the collective
flow. This hydrodynamic expansion boosts the momenta of the
pions, affecting both the numerator and the denominator
in $\langle (\Delta p_T)^2\rangle/\langle p_T\rangle^2$.
Although we do expect that the effect cancels in the 
ratio to a significant extent, the 
``blue shift'' approximation which we have used
is too simple. We have shown quantitatively
that the {\it fluctuations} in the flow velocity {\it can}
be neglected. However, the effects of the flow itself
are not sufficiently accurately treated as a uniform blue shift,
and must be treated more quantitatively in the future.
We find that our prediction
for $\langle (\Delta p_T)^2\rangle/\langle p_T\rangle^2$
is about $90\%$ of that which NA49 observes.
This gives us further
confidence that we can use thermodynamics to
understand the great bulk of the observed fluctuations; improving
the precision of the prediction 
by improving upon the blue shift approximation
remains to be done.

The data are precise enough that we can do more
than analyze the ``bulk'' of the fluctuations.
We can ask, for example, about the ratio $\sqrt{F}$
of $\langle N \rangle^{1/2} v_{\rm ebe}(p_T)$
to the variance of the inclusive single particle distribution.
This ratio is insensitive to the effects of the flow 
velocity.
For a classical ideal gas, $\sqrt{F}=1$.
We estimate that Bose
effects result in $\sqrt{F}=\sqrt{F_B}\approx 1.02$.
In the data, however, $\sqrt{F}=1.002\pm 0.002$.
The Bose effects may be small, but they are ten times
larger than the statistical error in the data.
The Bose correlations are being compensated by some
anti-correlation, and in Section 4 we find a possible
explanation.  

We show that energy conservation results in an anti-correlation
which is reduced by thermal contact between the direct
pions and an unobserved heat bath. The anti-correlation
vanishes if the heat bath has infinite heat capacity.
This effect, and indeed everything about the fluctuations 
we analyze, can be derived from the correlator $\langle \Delta n_p
\Delta n_k \rangle$ between the fluctuations of the occupation
numbers of pion modes with momenta $p$ and $k$. Energy conservation
implies that if $n_p$ fluctuates up, then
$n_k$ is more likely to fluctuate down. 
The magnitude
of the effect depends on the heat capacity of the
``heat bath'', but we estimate that 
it leads to $\sqrt{F} = \sqrt{F_B F_T}$ with 
$\sqrt{F_T}\approx 0.99$. 

With more detailed experimental study, either now
at the SPS, or soon at RHIC (STAR will study 
event-by-event fluctuations in $p_T$, $N$, particle ratios, etc;
PHENIX and PHOBOS in $N$ only) it should be possible 
to disentangle the different effects we describe.
Making a cut to look at only low $p_T$ pions should increase the effects of 
Bose enhancement. Bose enhancement effects are sensitive to $\mu_\pi$, 
and measuring these effects   
would allow one to make an experimental
determination of this quantity.  The anti-correlation
introduced by energy conservation and 
thermal contact 
is due
to terms in $\langle \Delta n_p\Delta n_k \rangle$
which are off-diagonal in $pk$. Thus, a direct measurement
of $\langle \Delta n_p\Delta n_k \rangle$ would make it
easy to separate this anti-correlation from other effects.
The cross correlation $\langle \Delta N \Delta p_T\rangle$
is a very interesting observable to study because it only
receives contributions from interesting effects, like
Bose enhancement, thermal contact and the critical fluctuations
discussed in Section 5. 
We hope that the combination of the theoretical tools we
have provided and the present NA49 data provide a solid
foundation for the future study of the thermodynamics of
the hadronic matter present at freeze-out in heavy ion collisions.

We also consider fluctuations in multiplicity $N$, an extensive observable.
These are not affected by
the boost which the pion momenta receive from the collective flow,
and this makes them easier to calculate than the fluctuations in $p_T$.
However, multiplicity fluctuations are contaminated 
experimentally by
fluctuations in the initial state, for example due to the distribution of
impact parameters.  This experimental
contamination can be reduced by 
making a tight enough centrality cut 
using a zero degree calorimeter.
We compare 
the multiplicity fluctuations of the $5\%$ most central events
in the NA49 data to those we predict from
a resonance gas, and find evidence that 
about 75\% 
of the observed
fluctuation is indeed thermodynamic in origin. 
We find that resonances play a
significant role in this comparison, increasing the magnitude
of thermodynamic fluctuations of the pion multiplicity and bringing
it closer to the data.

With the foundations established, we then describe
how the fluctuations we analyze will change if control
parameters are varied in
such a way that the baryon chemical potential  
at freeze-out, $\mu_{\rm f}$, moves toward and then past the critical
point in the QCD phase diagram at which a line 
of first order transitions ends at a second order endpoint.
We provide quantitative estimates of the magnitude of the
change in the observables we have analyzed which can
be expected 
near this point.
The agreement
between the noncritical thermodynamic fluctuations in $p_T$ which
we analyze in Section 3 and NA49 data make it unlikely
that central PbPb collisions at 160 AGeV freeze out
near the critical point.
Estimates we have made in a previous
paper suggest that the critical point is
located at a baryon chemical potential 
$\mu$ such that it will be found at
an energy between 160 AGeV and AGS energies. This makes it
a prime target for detailed study at the CERN SPS
by comparing data taken at 40 AGeV, 160 AGeV, and in between.
If the critical
point is located at such a low $\mu$ that the maximum
SPS energy is insufficient to reach it, 
it would
then be in a regime accessible to study by the 
RHIC experiments.  We want to stress that
we are more confident
in our ability to describe the properties of the
critical point and thus to predict {\it how} to find it than
we are in our ability to predict where it is.


The critical fluctuations near the endpoint 
affect the event-by-event fluctuations which we analyze in two different ways.
First, all effects of energy conservation 
should be greatly
reduced by thermal contact
as the critical fluctuations in the sigma field
cause the heat capacity to grow. Second, these critical
fluctuations have direct effects on the fluctuations
of the pions through the $G\sigma\pi\pi$ coupling.
We analyze the most singular effects of this coupling,
which are due to the zero momentum mode of the sigma field.
It is possible to analyze subleading corrections using
a diagrammatic approach, but we leave this to the future.

In the chiral limit, the critical point becomes
a tricritical point at which $G$ vanishes.  
We estimate the 
vacuum value of $G$ and use scaling laws
valid near a tricritical point to estimate $G$ at
the critical point. We then estimate the increase
in the fluctuations of $N$ and $p_T$ distributions
which we expect in heavy ion collisions which
freeze out near the critical point. Finite size
and finite time effects prevent $\langle (\Delta N)^2\rangle/\langle N\rangle$
and $\langle (\Delta p_T)^2\rangle\langle N\rangle$ from diverging,
as they would in an infinite system.  We estimate
that $\langle (\Delta N)^2\rangle/\langle N\rangle$ can grow by more
than 10\%. The ratio $\sqrt{F}$ which describes
the $p_T$-fluctuations becomes $\sqrt{F_B F_T F_\sigma}$
with $\sqrt{F_\sigma}$ about 1.1. This effect is not large
but is still predicted to be 
50 times larger than the statistical error in the present
NA49 measurement of $\sqrt{F}=1.002\pm 0.002$. 
We quantify the
uncertainty in our estimates in terms of the sigma
correlation length $\xi$ and the coupling $G$ at freezeout;
measurement of the enhanced fluctuations of $N$ and $p_T$ 
would allow one to estimate $G\xi$.

We want to emphasize that the ratio $\sqrt{F}$ is not the only observable
which can be used to detect the proximity of the critical point, and
indeed is not the most sensitive observable available.  We have
focussed on $\sqrt{F}$ because it is simple to define, and because NA49 has
published data to which we can compare our predictions.  However, the
specific form of the singularity in $\langle\Delta n_p\Delta
n_k\rangle$ which we find in Eq.~(\ref{dndnsigma}) tells us how to
construct observables which are more sensitive to the critical
fluctuations.  One possibility is to consider observables which are
sensitive to the off-diagonal part of $\langle\Delta n_p\Delta
n_k\rangle$, because the noncritical
off-diagonal anticorrelation in $\langle\Delta n_p\Delta
n_k\rangle$ should be replaced by a much
larger off-diagonal correlation near the critical point.  A second
possibility is an analysis of the cross correlation 
$\langle \Delta N \Delta p_T\rangle$.  
Because this cross correlation is dominated by interesting
effects, we have seen that it can increase by an order of magnitude at
the critical point.  A third possibility is to construct a ratio like
$\sqrt{F}$, but using only soft pions, with $p_T$ less than a specified cutoff.
The effects of the critical fluctuations are largest on the softest
pions, and they are therefore masked in $\sqrt{F}$ which receives significant
contribution from harder particles. Whereas we have found that the critical
fluctuations change $\sqrt{F}$ at the 10\% level, their effect on a ``soft
$\sqrt{F}\,$''  can easily be at the factor of two level.

Although the sigma quasiparticles themselves cannot be
reconstructed, their presence can be detected even more
directly than via their influence on the pions at freeze-out.
If freeze-out occurs near the critical point, some time
after freeze-out the sigma mass rises above the two pion
threshold, and the sigmas decay quickly. Since these pions
do not rethermalize, the resulting excess in the low-$p_T$
region of the pion momentum spectrum should be observable.
The mean $p_T$ of these pions is about $m_\pi/2$, and
they are almost as numerous as the direct pions. 
The event-by-event fluctuations in the multiplicity
of these soft pions would be even larger than those of the
rest of the pions near the critical point.  

In summary, our understanding of the thermodynamics of
QCD will be greatly enhanced by the detailed
study of event-by-event fluctuations in heavy
ion collisions.  We have estimated the influence of a number
of different physical effects, some special to the
vicinity of the critical point but many not, on
the fundamental correlator $\langle \Delta n_p\Delta n_k \rangle$.
This is itself measurable, but we have in addition used it
to make predictions for 
the fluctuations of observables which have been
measured at present, like
$\langle (\Delta p_T)^2\rangle$ and $\langle (\Delta N)^2\rangle$
and also for the cross correlation $\langle \Delta N\Delta p_T\rangle$.
The predictions of a simple resonance gas model, which
does not include critical fluctuations, is 
to this point in broad agreement
with the data. More detailed study, for example with varying
cuts in addition to new observables, will help to further constrain 
the nonthermodynamic fluctuations, which are clearly small,
and better understand the different thermodynamic effects.
The signatures we analyze allow experiments
to map out distinctive features of the QCD phase diagram.
The striking example which we have considered in
detail is the effect of a second order critical end point.
The nonmonotonic
appearance and then disappearance of any one of the signatures 
of the critical fluctuations which we have described 
would be strong evidence for the critical point. 
Furthermore, 
if a nonmonotonic
variation is seen in several of these observables, then
the maxima in all the 
signatures must occur simultaneously, at the same value
of the control parameters. Simultaneous detection of the
effects of the critical fluctuations on different observables
would turn strong evidence into an unambiguous discovery.

%% file: appendix.tex
\vspace{3ex}
{\bf Acknowledgements}

We are grateful to G. Roland for providing us with
preliminary NA49 data.
We acknowledge helpful conversations with 
M. Creutz, U. Heinz, M. Ga\'zdzicki, V. Koch,
St. Mr\'owczy\'nski, G. Roland and T. Trainor.

The work of MS is supported by NSF grant PHY97-22101.
The work of KR is supported in part by a DOE 
Outstanding Junior Investigator Award, by 
the Alfred P. Sloan
Foundation,  and by the DOE
under cooperative research agreement DE-FC02-94ER40818.
The work of ES is supported in part by DOE grant DE-FG02-88ER40388.

\section*{Note added in proof}

As we have stressed in Section 3, the fluctuations
in an extensive quantity such as the observed multiplicity
are unlike fluctuations in intensive quantities in that
they receive significant contributions from both (i) thermodynamic
fluctuations at freeze-out and (ii) nonthermodynamic fluctuations
during the initial stage of the collision.
Our approach has been to use a comparison
between the data and thermodynamic
predictions to constrain the magnitude of
non-thermodynamic fluctuations.

After this paper was submitted, Ref. \cite{BH} appeared. These 
authors have attempted a theoretical treatment
of those non-thermodynamic fluctuations
which are purely geometrical in that they can be attributed
to the distribution of impact parameters.  Further analysis
is presented in Ref. \cite{DS}.  These authors
include in addition the effects of fluctuations in the
NN cross-section \cite{Baym_etal},
which they find to be small, and also fluctuations in the number of 
punch-through spectators and effects due to the diffuse edges
of the incident nuclei, both of which are significant.
Combining all contributions to the multiplicity fluctuations,
thermodynamic and nonthermodynamic, yields fluctuations
which, with no new free parameters, reproduce the magnitude 
of the observed multiplicity fluctuations to 
within a few percent accuracy \cite{DS}.

\appendix

\section{Finiteness of Multiplicity}

Throughout the body of the paper,
we use event-by-event and inclusive averages defined 
probabilistically.  If we were interested in 
an infinite ensemble in which each member of the ensemble
was in the thermodynamic limit, no translation would be required.
However, 
when we want to compare the relations
involving quantities which are defined probabilistically
to those measured in an experiment, as in any application of probability
theory  we must have 
estimators for these quantities which can be constructed from
{\em finite} samples. In this Appendix, we discuss the effects 
due to finite sample size.

The typical size of these effects is one over a power of the sample size.
The total number, $C$, of events can be easily made very large
(say, $10^{6}$), so that $1/C$, and even $1/\sqrt C$ 
is much smaller than the {\em physical}
effects we consider, (such as Bose enhancement, for example) which are of the
order of a few percent or more. 
However, the number of pions in an event, $N$, is
limited by the size of the colliding system and 
the experimental acceptance of a detector, and is typically
of the order of a few hundred. This can introduce corrections of the
order of a fraction of a percent. Of course, these effects
are negligible when compared to the bulk of the fluctuations,
which we analyze in Section 3.
They are also smaller than the
effects we discuss in Sections 4 and 5, where we are interested
in signatures which rise and fall by of order 10\% near the
critical point.  
However, we have seen in Section 3 that the statistical 
errors in the present data are small enough that one
can compare
quantities like ${\langle N\rangle}v_{\rm ebe}^2$ and
$v_{\rm inc}^2$ to a precision of less than a percent. 
At this level, we must understand how to deal with 
the $1/N$ corrections.

Let us consider a sample of values of some one-particle observable
$q$. This sample
is broken into $C$ subsets, i.e., events, with $N_a$ values per event.
We use the notation: $q^a_i$, where $a=1\ldots C$ and $i=1\ldots N_a$.
(For example, $q^a_i$ may be the momentum of the $i$'th pion
in the $a$'th event.)
The numbers $q^a_i$ are distributed according to some (joint)
probability distribution. We assume that the expectation value
is the same for all $q^a_i$: $M[q^a_i]=m$. 

In Section 2 we pointed out that the inclusive mean
$\overline{q}^{\rm inc}$ is the same as the event-by-event average
$\langle q \rangle$. Both quantities are defined in the infinite $C$
and $N$ limit. Let us now try to estimate these two quantities, 
using our finite sample $q^a_i$. The natural estimate for
$\overline{q}^{\rm inc}$ is the following:
\begin{equation}\label{qinc}
\overline{q}^{\rm inc}_{\rm est} = {\sum_{a=1}^C\sum_{i=1}^{N_a} q^a_i
\over \sum_{a=1}^C N_a} = {1\over C\langle N \rangle} 
\sum_{a=1}^C\sum_{i=1}^{N_a} q^a_i
\end{equation}
where we have introduced (somewhat inconsistently, but suggestively):
\begin{equation}
\langle N\rangle = \frac1C \sum_{a=1}^C N_a,
\end{equation}
which is (an estimate for) the mean multiplicity in an event.
The total number of $q$'s in the sample is
$C\langle N\rangle$.
The property of the estimate (\ref{qinc}) is that its expectation
value is equal to $m$:
\begin{equation}
M[\overline{q}^{\rm inc}_{\rm est}] = m,
\end{equation}
for {\em any} $N$ or $C$. The standard
deviation of this quantity is $O(1/\sqrt{C\langle N\rangle})$, so
the estimate becomes perfect in the infinite $C$, $N$ limit.
Now let us estimate the event-by-event mean of $q$. An
estimate which appears natural is:
\begin{equation}\label{est0}
\langle q \rangle_{\rm est0} = \frac1C \sum_{a=1}^{C} \left(
{1\over N_a} \sum_{i=1}^{N_a} q^a_i\right).
\end{equation}
The expectation value of this estimate is:
\begin{equation}
M[\langle q \rangle_{\rm est0}] = m,
\end{equation}
and this estimate also becomes perfect as $C$, $N$ go to infinity.
However, 
\begin{equation}
\langle q \rangle_{\rm est0}\ne\overline{q}^{\rm inc}_{\rm est}.
\end{equation}
One can show that the difference between the two is on the
order of $\langle (\Delta N)^2\rangle/N^2 \sim 1/N$. It is obvious
how to improve the estimate (\ref{est0}) to make the relationship
$\langle q\rangle = \overline{q}^{\rm inc}$ hold {\em exactly}
for finite $N$. Writing:
\begin{equation}
\overline{q}^{\rm inc}_{\rm est} =
\frac1C \sum_{a=1}^{C} 
{N_a\over\langle N\rangle} 
\left({1\over N_a} \sum_{i=1}^{N_a} q^a_i\right) \stackrel{\rm def}= 
\langle q \rangle_{\rm est},
\end{equation}
we can interpret this definition of the estimate $\langle q
\rangle_{\rm est}$ as a result of averaging over events with
a weight proportional to the multiplicity in this event, $N_a$.
It is also clear intuitively that such a procedure is more
natural than taking all events with equal weight as is done in
(\ref{est0}). What is important is that, by construction, this procedure
rids us of any $1/N$ correction to the equality between
$\langle q\rangle_{\rm est}$ and $\overline{q}^{\rm inc}_{\rm est}$.

Let us now consider estimating variances of $q$. The natural
estimate for the inclusive square variance is:
\begin{equation}
v^2_{\rm inc}(q)_{\rm est} =  {1\over C\langle N \rangle - 1} 
\sum_{a=1}^C\sum_{i=1}^{N_a} (q^a_i - \overline{q}^{\rm inc}_{\rm
est})^2.
\end{equation}
Assuming that the variables $q^a_i$ are uncorrelated and
their dispersions are equal, i.e.:
\begin{equation}\label{mqq}
M[q^a_i q^b_j] = m^2 + \sigma^2\delta_{ij}\delta^{ab},
\end{equation}
one can show that
\begin{equation}
M[v^2_{\rm inc}(q)_{\rm est}] = \sigma^2,
\end{equation}
for any $C$, or $N$.
This is the property which we require of this estimate.

Next, we consider estimating the event-by-event variance. One seemingly 
natural candidate is:
\begin{equation}\label{vest0}
v^2_{\rm ebe}(q)_{\rm est0} = {1\over C - 1} \sum_{a=1}^C \left(
{1\over N_a} \sum_{i=1}^{N_a} q^a_i - \langle q \rangle_{\rm est}\right)^2.
\end{equation}
This is the procedure used by NA49 to calculate 
$v_{\rm ebe}(p_T)$ from their data, leading to the result shown
in
Table~\ref{tab:na49}. Let us calculate the expectation value
of this quantity, assuming that all $q$'s are independent as
in (\ref{mqq}). We find:
\begin{equation}\label{mest0}
M[v^2_{\rm ebe}(q)_{\rm est0}] = \sigma^2 {C\over C-1} \left(
\Big\langle \frac1N\Big\rangle - {1\over C\langle N \rangle} \right) \approx
\sigma^2 \Big\langle \frac1N \Big\rangle,
\end{equation}
where in the last approximate equality we neglected terms of relative
size ${\cal O}(1/C)$. In the case of completely uncorrelated $q$
we expect the following relation to hold between the $v^2_{\rm
ebe}(q)$ and $v^2_{\rm inc}(q)$:
\begin{equation}
\langle N\rangle v^2_{\rm ebe}(q) = v^2_{\rm inc}(q).
\end{equation}
This equality cannot and should not be satisfied for an arbitrary sample
(unlike the equality $\langle q\rangle = \overline{q}^{\rm inc}$),
but we want it to be satisfied on the level of expectation values:
\begin{equation}\label{mebe=minc}
\langle N\rangle M[v^2_{\rm ebe}(q)_{\rm est0}] = 
M[v^2_{\rm inc}(q)_{\rm est}].
\end{equation}
We see that for the estimate (\ref{vest0}) the difference between the
l.h.s. and the r.h.s of eq.(\ref{mebe=minc}) is
\begin{equation}\label{vebe0-vinc}
\langle N\rangle M[v^2_{\rm ebe}(q)_{\rm est0}] - M[v^2_{\rm inc}(q)_{\rm est}]
\approx \sigma \left(\langle N\rangle \Big\langle \frac1N\Big\rangle - 1\right)
\approx \sigma^2 {\langle(\Delta N)^2\rangle\over \langle N\rangle^2},  
\end{equation}
where we have neglected the 
${\cal O}(1/C)$ corrections and the corrections which 
are higher order in $1/N$.
The difference (\ref{vebe0-vinc})
is a $1/N$ effect, of course, but with $N\sim 300$ it
could easily reach a fraction of a percent. Note that we are
not talking here about statistical fluctuations around the mean
values which make the two estimates deviate from sample to sample.
Such effects are of the order $1/\sqrt C$ and can easily be made
negligible with sufficient statistics. Equation
(\ref{vebe0-vinc}), on the other hand,
reflects a {\em systematic} 
discrepancy between the expectation values of the estimates, which would only
go away if $N$, the number of particles 
in one event, were infinite.

Our task now is to give an estimator for $v^2_{\rm ebe}(q)$ which
satisfies (\ref{mebe=minc}) with no errors of order $1/N$
in the case when the  $q^a_i$ are uncorrelated. 
The lesson we learned from the
estimator for $\langle q\rangle$ suggests that we  take each event
with the weight $N_a/\langle N \rangle$. This gives:
\begin{equation}\label{vest}
v^2_{\rm ebe}(q)_{\rm est} = {1\over C - 1} \sum_{a=1}^C
{N_a\over\langle N\rangle}\left(
{1\over N_a} \sum_{i=1}^{N_a} q^a_i - \langle q \rangle_{\rm est}\right)^2.
\end{equation}
Calculating the expectation value we find that
\begin{equation}
M[v^2_{\rm ebe}(q)_{\rm est}] = {\sigma^2\over\langle N\rangle}
\end{equation}
exactly! This means that the estimate (\ref{vest}) satisfies
our criterion (\ref{mebe=minc}) exactly, to all orders in $1/N$
and $1/C$. This is the estimate for $v^2_{\rm ebe}(q)$ that 
should be used to analyze experimental data. It introduces
no $1/N$ or $1/C$ errors in the statement that the ratio $F=1$
in the absence of correlation or interaction between the pions.

We can avoid having to apply the
formula (\ref{vest}) to the original experimental data set
in order to 
recalculate the $v_{\rm
ebe}(p_T)$ 
given in Table~\ref{tab:na49}, which used the estimate
(\ref{vest0}). Using our analysis, we can instead
just use the fact that (the expectation values of) the two estimates
are related by (see (\ref{vebe0-vinc})):
\begin{equation}\label{estconv}
v_{\rm ebe}(q)_{\rm est} = v_{\rm ebe}(q)_{\rm est0} 
\left(1 
- \frac{1}{2}{\langle(\Delta N)^2\rangle\over \langle N\rangle^2}\right),
\end{equation}
up to corrections which are higher order in $1/N$ (and corrections of order
$1/\sqrt C$). We use this relation to convert from one estimate to
another in Eq. (\ref{vebeconversion}) of Section 3.4.

To finish the discussion of the $1/N$ effects we also point out
that yet another estimate for $v_{\rm ebe}(q)$ is used implicitly in the
definition of $\Phi_{p_T}$ in \cite{GM92,mrow3}:
\begin{equation}
\Phi_{p_T} = \langle N\rangle^{1/2} v_{\rm ebe}(p_T)_{\rm est\Phi} 
- v_{\rm inc}(p_T).
\end{equation}
The definition of $v_{\rm ebe}(p_T)_{\rm est\Phi}$
corresponds, in the language that we use here, to giving each event
a weight $N_a^2/\langle N \rangle^2$ (and using $C$ instead of $C-1$):
\begin{equation}
v^2_{\rm ebe}(q)_{\rm est\Phi} =
{1\over C} \sum_{a=1}^C
{N_a^2\over\langle N\rangle^2}\left(
{1\over N_a} \sum_{i=1}^{N_a} q^a_i - \langle q \rangle_{\rm est}\right)^2.
\end{equation}
Calculating the expectation value of this estimate, one finds:
\begin{equation}
M[v^2_{\rm ebe}(q)_{\rm est\Phi}] = \sigma^2 \left(
{1\over\langle N\rangle} - {\langle N^2\rangle\over C\langle N\rangle^3}
\right) \approx {\sigma^2\over\langle N\rangle},
\end{equation}
where we neglected ${\cal O}(1/C)$ corrections in the last step. We see that
in the $C\rightarrow\infty$ limit
this estimate does not suffer from $1/N$ corrections as far as
the relation (\ref{mebe=minc}) is concerned, and does not
differ from the estimate (\ref{vest}). 
However, it does
introduce small $1/C$ corrections, while the estimate (\ref{vest})
satisfies (\ref{mebe=minc}) exactly.